\documentclass[final,3p,authoryear,times]{elsarticle}
\usepackage{graphicx}
\usepackage{amsmath}

\usepackage{amssymb}
\usepackage{lineno}
\DeclareMathOperator{\sech}{sech}

\newcommand{\Alfven}{Alfv\'en }
\renewcommand{\div}{\mathbf{\nabla} \cdot}
\newcommand{\rot}{\mathbf{\nabla}\times}
\newcommand{\rmR}{\rm R}
\newcommand{\rmL}{\rm L}
\newcommand{\alphap}{\alpha^{+}}
\newcommand{\alpham}{\alpha^{-}}
\newcommand{\alphaxp}{\alpha_{x}^{+}}
\newcommand{\alphaxm}{\alpha_{x}^{-}}
\newcommand{\alphayp}{\alpha_{y}^{+}}
\newcommand{\alphaym}{\alpha_{y}^{-}}
\newcommand{\efd}{{\mathcal E}}

\newcommand*\patchAmsMathEnvironmentForLineno[1]{%
  \expandafter\let\csname old#1\expandafter\endcsname\csname #1\endcsname
  \expandafter\let\csname oldend#1\expandafter\endcsname\csname end#1\endcsname
  \renewenvironment{#1}%
     {\linenomath\csname old#1\endcsname}%
     {\csname oldend#1\endcsname\endlinenomath}}%
\newcommand*\patchBothAmsMathEnvironmentsForLineno[1]{%
  \patchAmsMathEnvironmentForLineno{#1}%
  \patchAmsMathEnvironmentForLineno{#1*}}%
\AtBeginDocument{%
\patchBothAmsMathEnvironmentsForLineno{equation}%
\patchBothAmsMathEnvironmentsForLineno{align}%
\patchBothAmsMathEnvironmentsForLineno{flalign}%
\patchBothAmsMathEnvironmentsForLineno{alignat}%
\patchBothAmsMathEnvironmentsForLineno{gather}%
\patchBothAmsMathEnvironmentsForLineno{multline}%
}

\journal{Journal of Computational Physics}

\begin{document}

\begin{frontmatter}



\title{Divergence-free Approximate Riemann Solver for the Quasi-neutral
Two-fluid Plasma Model}

%\title{Divergence-free HLL Approximate Riemann Solver for Quasi-neutral
%Two-fluid Plasma Simulations}


\author{Takanobu Amano\corref{corresponding}} \ead{amano@eps.s.u-tokyo.ac.jp}
\cortext[corresponding]{Corresponding author}

\address{Department of Earth and Planetary Science, University of Tokyo,
Tokyo, 113-0033, Japan}

\begin{abstract}
A numerical method for the quasi-neutral two-fluid (QNTF) plasma model is
described. The basic equations are ion and electron fluid equations and the
Maxwell equations without displacement current. The neglect of displacement
current is consistent with the assumption of charge neutrality. Therefore,
Langmuir waves and electromagnetic waves are eliminated from the system, which
is in clear contrast to the fully electromagnetic two-fluid model. It thus
reduces to the ideal magnetohydrodynamic (MHD) equations in the long
wavelength limit, but the two-fluid effect appearing at ion and electron
inertial scales is fully taken into account. It is shown that the basic
equations may be rewritten in a form that has formally the same structure as
the MHD equations. The total mass, momentum, and energy are all written in the
conservative form. A new three-dimensional numerical simulation code has been
developed for the QNTF equations. The HLL (Harten-Lax-van Leer) approximate
Riemann solver combined with the upwind constrained transport (UCT) scheme is
applied. The method was originally developed for MHD
\citep{2004JCoPh.195...17L}, but works quite well for the present model as
well. The simulation code is able to capture sharp multidimensional
discontinuities as well as dispersive waves arising from the two-fluid effect
at small scales without producing $\div \mathbf{B}$ errors. It is well known
that conventional Hall-MHD codes often suffer a numerical stability issue
associated with short wavelength whistler waves. On the other hand, since
finite electron inertia introduces an upper bound to the phase speed of
whistler waves in the present model, our code is free from the issue even
without explicit dissipation terms or implicit time integration. Numerical
experiments have confirmed that there is no need to resolve characteristic
time scales such as plasma frequency or cyclotron frequency for numerical
stability. Consequently, the QNTF model offers a better alternative to the
Hall-MHD or fully electromagnetic two-fluid models.
\end{abstract}


\end{frontmatter}

%\linenumbers

\section{Introduction}
\label{intro}

Understanding of a rich variety of nonlinear phenomena in space,
astrophysical, and laboratory plasmas requires numerical simulations at
various levels of approximations. At the largest scale, magnetohydrodynamic
(MHD) description is useful because the scale-free nature of MHD allows us to
conduct simulations with a realistic scale size. On the other hand, physics at
kinetic scales (i.e., ion and electron inertial lengths) must be taken into
account in cases where it plays the key role. A well-known example is the
diffusion region of collisionless magnetic reconnection, in which the kinetic
effect is essential for violating the frozen-in condition. Fully kinetic
particle-in-cell (PIC) simulations have been used to investigate such
problems. It is important to point out that the characteristics of spatially
localized tiny regions may ultimately affect even the global dynamics of the
system.

Although it is believed that physics beyond MHD ultimately needs to be
incorporated properly even for the modeling of macroscopic phenomena, it is
still a formidable task, albeit not impossible, to perform fully kinetic PIC
simulations at a macroscopic scale. In practice, it is desirable to start with
a simpler model and gradually proceed toward better (but more complicated)
physics models with less approximations. Hall-MHD is one of such better models
in the sense that it takes into account physics at the ion inertial scale. The
hybrid simulation model that deals with kinetic ions and a massless fluid
electron can be considered as a ``kinetic version'' of Hall-MHD. The Hall-MHD
and hybrid models are therefore believed to be possible alternatives to MHD
for the next generation global modeling.

Although both Hall-MHD and hybrid have been well established standard models
for simulations of collisionless plasmas, it is well known that they often
suffer a numerical difficulty due to high frequency whistler waves. The
dispersion relation of whistler waves $\omega \propto k^2$ leads to the
increase in the phase speed at short wavelength without bound. This is
generally thought of as a source of numerical instability. A common strategy
to stabilize such simulations is to introduce ad-hoc numerical dissipation
such as hyper-resistivity in the code
\citep{2001JGR...106.3773M,2001JGR...106.3759S}. However, it is not easy to
control the amount of numerical dissipation with this kind of approach.
Furthermore, since the strategy is quite different from the philosophy of
modern high-order shock capturing schemes, this makes it difficult to extend
such codes to the Hall-MHD regime. Although one may use an implicit scheme to
circumvent the problem, this will make implementation of the algorithm much
more complex \citep[e.g.,][]{2008CoPhC.178..553A,2008JCoPh.227.6967T}.

A more straightforward approach is to employ the fully electromagnetic
two-fluid (EMTF) plasma model in which the full set of Maxwell equations are
coupled with two separate (i.e., ion and electron) fluids equations
\citep{2003JCoPh.187..620S,2005CoPhC.169..251L,2006JCoPh.219..418H,2011PhPl...18i2113S,
Kumar2012}. The phase speed of whistler waves has an upper bound in this
system due to the presence of finite electron inertia. On the other hand,
since it is essentially a fluid counterpart of the PIC simulation, it must
deal with high frequency Langmuir waves as well as electromagnetic
waves. Numerical stability requires that these waves should adequately be
resolved by the simulation time step unless more complicated implicit schemes
are employed \citep{Kumar2012}. In general, however, these high frequency
waves are not of interest as far as macroscopic dynamics is concerned. The
neglect of displacement current (which implies the charge neutrality) in the
Maxwell equations is indeed a reasonable assumption if one considers
non-relativistic problems, although this does not in general apply to highly
relativistic plasmas \citep[e.g.,][]{2013ApJ...770...18A}.

There is also a way to incorporate finite electron inertia effect into
Hall-MHD/hybrid without resorting to the full set of Maxwell equations.
Conventionally, finite electron inertia effect has been included as a
correction to the magnetic field
\citep[e.g.,][]{1998JGR...103..199K,1998JGR...103.9165S,2001JGR...106.3759S,
2008JGRA..113.9204N}. Although most of previous studies adopted some kind of
simplification, the finite electron inertia effect if appropriately included
can correctly introduce an upper bound to the phase speed of whistler
waves. On the other hand, the motivation for these studies was to initiate
spontaneous magnetic reconnection without relying on an anomalous resistivity
model. Therefore, a possible advantage of finite electron inertia effect on
the numerical stability issue has not been paid much attention. We have
recently shown that, by modifying the procedure to incorporate finite electron
inertia into the model, hybrid simulations can be made much more robust
particularly in low-density regions where whistlers become problematic
\citep{2014JCoPh.275..197A}. This was made possible by implementing finite
electron inertia as a correction to the electric field (i.e., the generalized
Ohm's law), which is then used to update the magnetic field. This physically
more consistent approach gives a natural way to handle even a pure vacuum
region in a hybrid code. It is quite natural to expect that essentially the
same methodology can be applied to Hall-MHD equations, because kinetic ion
dynamics does not play a role for dispersion of whistler waves.

In the present paper, we consider a system consisting of two-fluid equations
coupled with Maxwell equation without displacement current (Darwin
approximation), which we call the quasi-neutral two-fluid (QNTF) model. As we
will see in the next section, it is approximately the same as the Hall-MHD
equations with finite electron inertia, but terms dropped in previous studies
are retained for consistency. Consequently, the total mass, momentum, and
energy including finite electron contributions are all written in the
conservative form. The conservation laws coupled with the induction equation
for the magnetic field have the same formal structure as the MHD equations,
which thus may be solved by a known conservative scheme. Because of the neglect
of displacement current, there are no high frequency waves such as Langmuir or
electromagnetic waves, and the number of eigenmodes is indeed the same as
MHD. The system correctly reduces to the ideal MHD in the long wavelength
limit. Therefore, we think that it provides a natural extension of MHD having
desirable properties both in terms of physics and numerics.

We have developed a three-dimensional (3D) numerical simulation code to solve
the proposed system of equations. We employ the single-state HLL
(Harten-Lax-van Leer) approximate Riemann solver as a building block. It only
requires the maximum characteristic speed, and is independent of detailed
information on the eigenmode structure. The scheme is thus suitable to the
QNTF equations because eigenmode decomposition for this system should
certainly be much more laborious task than for MHD. In addition, we adopt the
Upwind Constrained Transport (UCT) scheme to keep the divergence error of the
magnetic field within machine accuracy \citep{2004JCoPh.195...17L}. The UCT
scheme is based on the Constrained Transport (CT) scheme
\citep{1988ApJ...332..659E}, but is designed specifically to be consistent
with an underlying Riemann solver. Although it was originally developed for
MHD, we found it is useful for the QNTF equations as well. With these
numerical techniques, our simulation code is able to capture sharp
discontinuities as well as dispersive waves arising from the two-fluid effect
at the same time even in multidimensions without violating the divergence-free
property.

In the next section, we introduce the QNTF model. The characteristics of the
model and its advantages over the Hall-MHD and EMTF models are
discussed. Section 3 is devoted to numerical algorithm used in our
code. Numerical results of several benchmark problems are discussed in section
4. Finally, conclusions are given in section 5.

\section{Quasi-neutral Two-fluid Model}
\subsection{Basic Equations}
\label{sec:basic_equations}
We start with the following fluid equations for a particle species $s$ ($i$
and $e$ for ion and electron respectively) of charge $q_s$ and mass $m_s$:
\begin{align}
 & \frac{\partial}{\partial t} \rho_s +
 \div \left( \rho_s \mathbf{v}_s \right) = 0 \\
 & \frac{\partial}{\partial t} \rho_s \mathbf{v}_s +
 \div
 \left( \rho_s \mathbf{v}_s \mathbf{v}_s + p_s \mathbf{I} \right) =
 \frac{q_s}{m_s} \rho_s
 \left( \mathbf{E} + \frac{\mathbf{v}_s}{c} \times \mathbf{B} \right) \\
 & \frac{\partial}{\partial t} \left( \frac{1}{2} \rho_s
 \mathbf{v}_s^2 + \frac{1}{\gamma-1} p_s \right) + \div \left\{ \left(
 \frac{1}{2} \rho_s \mathbf{v}_s^2 + \frac{\gamma}{\gamma-1} p_s \right)
 \mathbf{v}_s \right\} = \frac{q_s}{m_s} \rho_s \mathbf{v} \cdot \mathbf{E},
\end{align}
where $\rho_s$, $\mathbf{v}_s$, $p_s$ are the mass density, bulk velocity, and
(scalar) pressure (with $\mathbf{I}$ being the unit tensor), respectively. We
here assume a polytropic equation of state with a specific heat ratio denoted
by $\gamma$ (independent of particle species). The right-hand side of the
above equations represents the Lorentz force with the electromagnetic field
$\mathbf{E}$, $\mathbf{B}$, and the speed of light $c$.

Since we only consider low frequency phenomena, the displacement current in
the Maxwell equations is ignored.
\begin{align}
 & \frac{1}{c} \frac{\partial}{\partial t} \mathbf{B} = - \rot \mathbf{E}
 \label{eq:faraday} \\ & \rot \mathbf{B} = \frac{4 \pi}{c} \mathbf{J}.
 \label{eq:ampere}
\end{align}
As usual, the electric current density is given by a sum of contributions from
ions and electrons
\begin{align}
 \mathbf{J} = \frac{q_i}{m_i} \rho_i \mathbf{v}_i + \frac{q_e}{m_e} \rho_e
 \mathbf{v}_e \label{eq:current}
\end{align}
and we always assume charge neutrality
\begin{align}
 0 = \frac{q_i}{m_i} \rho_i + \frac{q_e}{m_e} \rho_e, \label{eq:charge}
\end{align}
which may also be written as $n_i = n_e$ for $q_i =-q_e = e$. Here, $n_s$ is
the number density and $e$ is the elementary charge. This is indeed consistent
with the neglect of the displacement current, because the longitudinal part of
the displacement current represents charge-density fluctuations. The
solenoidal condition for the magnetic field
\begin{align}
 \div \mathbf{B} = 0
\end{align}
gives a constraint that must be satisfied.

The above equations have source terms in the right-hand side and the energy
and momentum are not strictly conserved quantities in numerical simulations.
Instead, we use the following computationally more convenient conservative
form of equations that are obtained by taking sum of the two species:
\begin{align}
 \frac{\partial}{\partial t} \mathbf{U} + \div \mathbf{F} = 0,
 \label{eq:fluid}
\end{align}
where
\begin{align}
 \mathbf{U} =
 \begin{pmatrix}
  \displaystyle \rho_i + \rho_e
  \\
  \displaystyle \rho_i \mathbf{v}_i + \rho_e \mathbf{v}_e
  \\
  \displaystyle
  \frac{1}{2} \rho_i \mathbf{v}_i^2 +
  \frac{1}{2} \rho_e \mathbf{v}_e^2 +
  \frac{1}{\gamma-1} (p_i + p_e) +
  \frac{\mathbf{B}^2}{8 \pi}
 \end{pmatrix}
 \equiv
 \begin{pmatrix}
  D
  \\
  \mathbf{M}
  \\
  K
 \end{pmatrix}
\end{align}
and
\begin{align}
 \mathbf{F} =
 \begin{pmatrix}
  \displaystyle \rho_i \mathbf{v}_i + \rho_e \mathbf{v}_e
  \\
  \displaystyle
  \rho_i \mathbf{v}_i \mathbf{v}_i +
  \rho_e \mathbf{v}_e \mathbf{v}_e +
  \left( p_i + p_e + \frac{\mathbf{B}^2}{8 \pi} \right) \mathbf{I} -
  \frac{\mathbf{B} \mathbf{B}}{4 \pi}
  \\
  \displaystyle
  \left( \frac{1}{2} \rho_i \mathbf{v}_i^2 + \frac{\gamma}{\gamma-1} p_i \right)
  \mathbf{v}_i +
  \left( \frac{1}{2} \rho_e \mathbf{v}_e^2 + \frac{\gamma}{\gamma-1} p_e \right)
  \mathbf{v}_e + \frac{c}{4 \pi} \mathbf{E} \times \mathbf{B}
 \end{pmatrix}
\end{align}
represent conservative variables and their corresponding fluxes. Here
Eqs.~(\ref{eq:faraday}-\ref{eq:charge}) have been used to rewrite the Lorentz
force on the right-hand side into the above conservative form. Note that the
same strategy was recently used in fully relativistic two-fluid simulations
(i.e., relativistic version of EMTF), so that the total energy and momentum
become strictly conserved quantities \citep{2013ApJ...770...18A}.

As shown in \ref{sec:ohmslaw}, the generalized Ohm's law for the present
system may be written as
\begin{align}
 \left(\Lambda + c^2 \nabla \times \nabla \times \right) \mathbf{E} = -
 \frac{\mathbf{\Gamma}}{c} \times \mathbf{B} + \div \mathbf{\Pi} + \eta
 \Lambda \mathbf{J}.
 \label{eq:ohm}
\end{align}
Here we have introduced resistivity $\eta$ in a rather ad-hoc manner to take
into account phenomenological collisions, although it is absent ($\eta = 0$)
in the original ideal two-fluid equations. The moment quantities $\Lambda$,
$\mathbf{\Gamma}$, $\mathbf{\Pi}$ appearing in the above equation are defined
as follows
\begin{align}
 \Lambda &=
 4 \pi \rho_i \frac{q_i^2}{m_i^2} + 4 \pi \rho_e \frac{q_e^2}{m_e^2} =
 \omega_{pi}^2 + \omega_{pe}^2 \\
 \mathbf{\Gamma} &=
 4 \pi \rho_i \frac{q_i^2}{m_i^2} \mathbf{v}_i +
 4 \pi \rho_e \frac{q_e^2}{m_e^2} \mathbf{v}_e =
 \omega_{pi}^2 \mathbf{v}_i + \omega_{pe}^2 \mathbf{v}_e \\
 \mathbf{\Pi} &= \frac{4 \pi q_i}{m_i}
 \left( \rho_i \mathbf{v}_i \mathbf{v}_i + p_i \mathbf{I} \right) +
 \frac{4 \pi q_e}{m_e} \left( \rho_e \mathbf{v}_e \mathbf{v}_e + p_e
 \mathbf{I} \right),
\end{align}
where $\omega_{ps}^2 \equiv 4 \pi \rho_s q_s^2/m_s^2$ is the plasma frequency
for a particle species $s$. The connection between Eq.~(\ref{eq:ohm}) and
conventional Ohm's laws will be discussed in the next subsection. We should
emphasize that, aside from the resistivity $\eta$, this form of the
generalized Ohm's law is exact. It is obtained from the basic equations
without any approximations or assumptions. Notice that the above equation is
an implicit equation for the electric field. Therefore, in general, matrix
inversion is needed to obtain the electric field. Detail will be discussed
later in section \ref{sec:efield}.

Eqs. (\ref{eq:current}-\ref{eq:charge}) clearly indicate that the density and
velocity for each species are not independent. Although the electron density
and velocity appear frequently in this paper for clarity of notation, they
must always be replaced by
\begin{align}
 & \rho_e = - \rho_i \frac{q_i/m_i}{q_e/m_e}
 \label{eq:ele_density}\\
 & \mathbf{v}_e = \mathbf{v}_i -  \frac{m_i}{q_i} \frac{c}{4 \pi \rho_i}
 \rot \mathbf{B} \label{eq:ele_velocity}
\end{align}
in actual calculations.

In addition, the ion to electron temperature ratio denoted by $\tau \equiv
T_i/T_e$ is assumed to be a given constant throughout in this paper. This
implies that the energy exchange between the species occurs instantaneously
and the temperature ratio quickly relaxes to the prescribed constant value
$\tau$ (typically chosen to be $\tau = 1$). We adopt this simplification
because the energy distribution among different species through a dissipative
process (such as a shock wave) is not known a priori within the framework of a
fluid model. It crucially depends on complicated kinetic physics in an
unresolved dissipative layer. We note that there is no fundamental difficulty
in dealing with $T_i$ and $T_e$ independently without such a simplifying
assumption. Indeed, in such a case, we have found that there appears an
entropy-like mode which is a pressure-balanced structure across which only ion
and electron temperatures are exchanged with keeping the total gas pressure
(and other quantities) unchanged. However, in general, it is only the total
gas pressure behind the dissipative layer that we can correctly predict (from
the Rankine-Hugoniot relations), and the temperature ratio is likely to be
affected by numerical dissipation. Since the total gas pressure does not
depend on a particular choice of $\tau$, we have not observed any noticeable
differences between the simplified and the more rigorous implementations. We
thus think the assumption adopted here is a reasonable simplification.

In summary, the fluid quantities and magnetic field are updated respectively
by using Eq.~(\ref{eq:fluid}) and Eq.~(\ref{eq:faraday}). The electric field
appearing in these equations is determined by the generalized Ohm's law
Eq.~(\ref{eq:ohm}). These equations and the relationship
Eq.~(\ref{eq:current}-\ref{eq:charge}) with the constant ion and electron
temperature ratio $\tau$ close the system of equations, which is called the
QNTF model. It is important to mention that the number of eigenmodes in this
system is seven, which is the same as MHD. Namely, one may consider
$\mathbf{u} = \{\rho_i, \mathbf{v}_i, p_i, \mathbf{B}\}$ as primitive
variables (with the $\nabla \cdot \mathbf{B} = 0$ constraint). Given the ion
quantities and the magnetic field, the electron quantities are automatically
determined by Eq.~(\ref{eq:ele_density}-\ref{eq:ele_velocity}). As in the case
of MHD, the electric field is also essentially a dependent variable, since it
is completely specified by the primitive variables.

Obviously, the advantage of using the conservative form Eq.~(\ref{eq:fluid})
instead of the original two-fluid equations written separately is that the
exact conservation of total energy and momentum may be guaranteed if a
conservative scheme is used for numerical computation. Furthermore, since the
QNTF equations in the conservative form are very similar to MHD equations, one
may use numerical methods developed for MHD with relatively minor
modifications. In particular, the absence of the Lorentz force as the source
term gives an advantage, because otherwise it would possibly impose a
constraint on the time step for numerical stability. As we will demonstrate
with the numerical examples presented in section \ref{sec:results}, the time
step of our code is restricted by the fastest wave mode, and there is no need
to resolve characteristic time scales such as the cyclotron frequency.

For the sake of completeness, we here give explicit formulae to calculate the
primitive variables from the conservative variables. Given the conservative
variables $\mathbf{U} = \{D, \mathbf{M}, K\}$, one can calculate $\rho_i$ and
$\mathbf{v}_i$ according to
\begin{align}
 & \rho_i =
 \frac{q_e}{m_e} D /
 \left( \frac{q_e}{m_e} - \frac{q_i}{m_i} \right),
 \\
 & \mathbf{v}_i =
 \left(
 \frac{q_e}{m_e} \mathbf{M} - \frac{c}{4 \pi} \nabla \times \mathbf{B}
 \right) / \frac{q_e}{m_e} D.
\end{align}
The electron density and velocity are determined by using
Eqs.~(\ref{eq:ele_density}-\ref{eq:ele_velocity}), from which one may obtain
the ion pressure $p_i$ (and also electron pressure $p_e = p_i/\tau$) as follows
\begin{align}
 p_i = \frac{\gamma-1}{1 + 1/\tau}
 \left(
 K -
 \frac{1}{2} \rho_i \mathbf{v}_i^2 -
 \frac{1}{2} \rho_e \mathbf{v}_e^2 -
 \frac{\mathbf{B}^2}{8 \pi}
 \right).
\end{align}

\subsection{Model Characteristics}
\label{sec:model_characteristics}

It is helpful to introduce approximations to the generalized Ohm's law
Eq.~(\ref{eq:ohm}) in a step-by-step manner to see the relationship with more
familiar forms of Ohm's law. First, notice that $\Lambda$ and
$\mathbf{\Gamma}$ are the sum of contributions of the two species that are
proportional to the plasma frequency. Similarly, the contributions to
$\mathbf{\Pi}$ are inversely proportional to the mass. Therefore, one may
safely ignore the ion fluid contributions and adopt the following
approximation:
\begin{align}
 &\Lambda \approx \omega_{pe}^2,
 \\
 &\mathbf{\Gamma} \approx \omega_{pe}^2 \mathbf{v}_e,
 \\
 &\mathbf{\Pi} \approx \frac{4 \pi q_e}{m_e} \left( \rho_e \mathbf{v}_e
 \mathbf{v}_e + p_e \mathbf{I} \right),
\end{align}
for $m_e/m_i \ll 1$. This yields the following Ohm's law:
\begin{align}
 \left(1 + \frac{c^2}{\omega_{pe}^2} \nabla \times \nabla \times \right)
 \mathbf{E} = -
 \frac{\mathbf{v}_e}{c} \times \mathbf{B}
 - \frac{m_e}{\rho_e} \div
 \left(
 \mathbf{\rho_e \mathbf{v}_e \mathbf{v}_e + p_e \mathbf{I}}
 \right)
 + \eta \mathbf{J},
 \label{eq:ohm_approx}
\end{align}
which includes correction terms approximately describe the finite electron
inertia effect. Finite electron inertia codes that have been used in previous
studies, adopt essentially the same approximation (except for some minor
differences). Now, it is easy to understand that the second term in the
left-hand side of the above equation is smaller than the first term by a
factor $k^2 c^2/\omega_{pe}^2$, which thus can be ignored for scale length
longer than the electron inertial length $c/\omega_{pe}$. If we further drop
the convective derivative term ($\propto \rho_e \mathbf{v}_e \mathbf{v}_e$),
we obtain Ohm's law for the Hall-MHD regime (i.e., with a massless electron
fluid):
\begin{align}
 \mathbf{E} =
 - \frac{\mathbf{v}_i}{c} \times \mathbf{B}
 + \frac{m_i}{q_i} \frac{1}{4 \pi \rho_i} (\rot \mathbf{B}) \times \mathbf{B}
 - \frac{m_e}{\rho_e} \nabla p_e
 + \eta \mathbf{J},
 \label{eq:ohm_hall}
\end{align}
where the electron velocity has been replaced by
Eq.(\ref{eq:ele_velocity}). It is well known that we obtain Ohm's law for the
MHD regime by taking the long wavelength (longer than the ion inertial
length), and cold electron limit ($p_e \rightarrow 0$). This analysis confirms
that Eq.~(\ref{eq:ohm}) indeed generalizes the known forms of Ohm's law. From
the rigorous Ohm's law without approximations, we see that the magnetic field
convection (or frozen-in) velocity must be given by
\begin{align}
 \frac{\mathbf{\Gamma}}{\Lambda} =
 \frac{\omega_{pe}^2 \mathbf{v}_e + \omega_{pi}^2 \mathbf{v}_i}
 {\omega_{pi}^2 + \omega_{pe}^2}
 \approx \mathbf{v}_e
 + \frac{m_e}{m_i} \left( \mathbf{v}_i - \mathbf{v}_e \right)
 + O\left(\frac{m_e^2}{m_i^2}\right)
\end{align}
rather than the electron velocity $\mathbf{v}_e$. The finite electron inertia
correction appearing in the above equation has been ignored in previous
studies, although one may expect it to be a relatively minor correction.

There is another concern associated with the approximation
Eq.~(\ref{eq:ohm_approx}). Namely, it will break Galilean invariance due to
the appearance of an unphysical electric field. To see this, consider for
simplicity a cold ($p_i = p_e = 0$), current-free ($\rot \mathbf{B} = 0$) MHD
flow. From Eq.~(\ref{eq:ele_velocity}), it follows that the ion and electron
flow velocities must be the same $\mathbf{v}_i = \mathbf{v}_e =
\mathbf{V}$. In this case, the ion and electron contributions to ${\mathbf
\Pi}$ in the exact equation cancel with each other:
\begin{align}
 \mathbf{\Pi} =
 \left( \frac{4 \pi q_i}{m_i} \rho_i + \frac{4 \pi q_e}{m_e} \rho_e \right)
 \mathbf{V} \mathbf{V} = 0
\end{align}
because of the charge neutrality assumption Eq.~(\ref{eq:charge}). On the
other hand, it is obvious that the approximate expression of $\mathbf{\Pi}$
remains finite and produces an unphysical electric field when the flow speed
or density has spatial variation (i.e., $\div \mathbf{\Pi} \neq 0$). Although
the magnitude of the electric field will be small in comparisons with other
contributions unless the flow speed is highly supersonic, it is better to keep
the ion contributions included for consistency. The problem arises because
careless neglect of terms of order $O(m_e/m_i)$ breaks the local momentum
conservation law.

In the present model, the finite electron inertia effect is fully taken into
account in the sense that it is correct to all orders of $m_e/m_i$. This
enables us to write the equations for total energy and momentum including the
electron contributions in the conservative form. This property has been
missing in existing finite electron inertia codes. Consequently, the model is
valid even for a {\it pair plasma} $m_i = m_e$, although application of a
non-relativistic pair plasma model would be limited in practice. This may be
sometimes useful for a control experiment because in this case the Hall term
disappears and the perfect symmetry is preserved as in the case of MHD.

One may recognize the present model as a better alternative to the Hall-MHD
(with or without finite electron inertia) or EMTF models. In Hall-MHD,
dispersive whistler waves often pose numerical difficulty because the phase
speed increases without bound at short wavelength. By taking into account
finite electron inertia effect, there appears an upper bound in the phase
speed that may improve numerical stability (see \ref{sec:linear} for
detail). The fact that the basic equations are written in the conservative
form makes it easy to apply a known scheme for a hyperbolic conservation
law. As we will see in the numerical examples discussed in section
\ref{sec:results}, the QNTF model automatically reduces to MHD in the long
wavelength limit. This property allows us to use the same shock-capturing code
to investigate the dependence on the scale size of the problem without
introducing ad-hoc numerical stabilization techniques. This is in clear
contrast to the EMTF model where the displacement current is retained in the
Maxwell equation. In the EMTF model, there remain high frequency waves
(electromagnetic and Langmuir waves) even in the long wavelength limit. These
waves impose a severe restriction on the time step of an explicit time
integration scheme. We thus think, in situations where high frequency waves do
not play a major role, our model is better than the EMTF model in practice.

One may expect that the QNTF model gives a good
approximation to the EMTF model when the following condition is met
\begin{align}
 \frac{\Omega_{ce}^2}{\omega_{pe}^2} =
 \left( \frac{V_{A,e}}{c} \right)^2 \ll 1,
 \label{eq:condition}
\end{align}
where $\Omega_{ce} = q_e B / m_e c$ is the electron cyclotron frequency and
$V_{A,e} = B/\sqrt{4 \pi \rho_e}$ is the electron \Alfven speed ($\sim
\sqrt{m_i/m_e}$ times the \Alfven speed). In this case, the two time scales
are well separated and the interaction between them may be assumed to be
weak. The above condition thus implies that relatively slow (slower than the
electron cyclotron period) time scale phenomena being described by the QNTF
model are almost decoupled from higher frequency phenomena $\sim
\omega_{pe}$. It is also possible to estimate a normalized charge density
fluctuation amplitude using the Gauss law $\div \mathbf{E} = 4 \pi \rho_c$
(where $\rho_c$ is the charge density) as:
\begin{align}
 \left| \frac{n_i - n_e}{n_0} \right| \sim
 \left( \frac{c/\omega_{pe}}{L} \right)
 \left( \frac{V}{c} \right)
 \left( \frac{V_{A,e}}{c} \right),
\end{align}
where $L$ and $V$ are typical spatial scale length and velocity,
respectively. Note that the electric field strength is estimated by $E \sim V
B/c$. If one takes $L \sim c/\omega_{pe}$ and $V \sim V_{A,e}$, the right-hand
side becomes $\Omega_{ce}^2/\omega_{pe}^2$. This again suggests that the QNTF
model is appropriate when Eq.~(\ref{eq:condition}) is satisfied.

From this analysis, we confirm that it is reasonable to neglect the
displacement current for modeling low frequency non-relativistic plasma
phenomena (i.e., $V_{A,e}/c \ll 1$). Strictly speaking, however, it does not
prove the validity of our assumption, which must be tested ultimately by
direct comparison between the QNTF and EMTF models. This will be addressed in
a future publication.

We note that, even if the condition Eq.~(\ref{eq:condition}) is satisfied,
application of the present model to phenomena of scale size less than the
electron inertial length should be done with care. This is because electron
kinetic effect is not properly taken into account in a fluid model, which
will, however, play a role at this scale unless the electron fluid is
unusually cold. Nevertheless, the inclusion of finite electron inertia is of
critical importance at least for numerical stability even in the absence of
kinetic effect.

\subsection{Effect of Resistivity}
\label{sec:resistivity}

In analogy with MHD, it is easy to understand that the parameter $\eta$
appearing in the generalized Ohm's law Eq.~(\ref{eq:ohm}) represents the
resistivity in the usual sense. However, at scales comparable to or less than
the electron inertial length, it does not lead to magnetic diffusion.

To demonstrate this, consider for simplicity a resistive medium so that the
left-hand side of Eq.~(\ref{eq:ohm}) balances with the resistive term. For a
long wavelength mode $k c/\omega_{pe} \ll 1$, the induction equation for the
magnetic field becomes a diffusion equation
\begin{align}
 \frac{\partial}{\partial t} \mathbf{B} \approx \frac{\eta c^2}{4 \pi}
 \nabla^2 \mathbf{B},
\end{align}
from which one can immediately understand that $\eta$ is actually proportional
to the magnetic diffusivity. This is consistent with the usual resistive MHD
equations. For $k c/\omega_{pe} \gg 1$, on the other hand, Ohm's law becomes
\begin{align}
 c^2 \nabla \times \nabla \times \mathbf{E} \approx \eta \Lambda \mathbf{J}.
\end{align}
In this case, we obtain the following equation by taking rotation of the
induction equation
\begin{align}
 \frac{\partial}{\partial t} \mathbf{J} \approx - \eta \frac{\Lambda}{4 \pi}
 \mathbf{J}.
\end{align}
This is a pure damping equation that does not involve any spatial derivatives.
In contrast to the diffusion equation, the electric current is damped locally
without propagating to neighboring regions in this parameter
regime. Therefore, the restriction on time step required for numerical
stability, which would be severe for a diffusion equation in a strongly
resistive medium, is very much relaxed. This form of resistivity was recently
used in \cite{2014JCoPh.275..197A} to improve the stability of a hybrid code
in and around a vacuum region.

This difference may be understood as follows. Physically, a finite $\eta$
arises if there is a friction between the two fluids. The friction may be
either due to Coulomb collisions or wave-particle interactions associated with
unresolved microscopic turbulence (i.e., anomalous resistivity). For scales
less than the electron inertial length, neither electrons nor ions are
frozen-in to the magnetic field line. Therefore, the collision does not
directly alter the magnetic field evolution. On the other hand, the relative
streaming between the two fluids decays exponentially due to the friction, so
does the electric current. We should mention that although the resistivity
works differently in the two different parameter regimes, in both cases, the
system relaxes to the same current-free state $\rot \mathbf{B} = 0$.

In the following, the normalized resistivity defined as
\begin{align}
 \eta^{\star} \equiv \frac{\omega_p}{4 \pi} \eta
\end{align}
is used. Here $\omega_p^2 = \Lambda = \sum_{s} \omega_{ps}^2$ is defined with
the local density, and thus, the collisionality is assumed to depend on the
local plasma frequency. This is reasonable for modeling anomalous resistivity
as its effective collision frequency will be characterized more or less by the
plasma frequency.

\section{Numerical Algorithm}
We now describe the numerical algorithm used in our newly developed 3D
simulation code. As has already become clear in the previous section, the QNTF
equations consist of two coupled subsystems: conservation laws for five scalar
(conservative) variables Eq.~(\ref{eq:fluid}) and the induction equation for
the magnetic field Eq.~(\ref{eq:faraday}). Numerical solutions must satisfy
the divergence-free condition $\div \mathbf{B} = 0$ as much as possible so as
to minimize the loss of accuracy in multidimensions
\citep[e.g.,][]{1999JCoPh.149..270B,2000JCoPh.161..605T}.

Because of the formal similarity with the MHD equations, we can apply some of
numerical methods developed for MHD. Here we adopt the HLL approximate Riemann
solver combined with the UCT scheme
\citep[HLL-UCT;][]{2000ApJ...530..508L,2004JCoPh.195...17L}, that satisfies
the divergence-free condition up to machine accuracy. The scheme was
originally developed for classical MHD, and successfully applied to relativistic
MHD as well \citep{2003A&A...400..397D,2007A&A...473...11D}. Although the
generalized Ohm's law used in this study differs considerably from the ideal
MHD, we find that the concept of the UCT is still useful for the system
considered in this paper.

Below we discuss only approximation in space; i.e., the temporal derivative
always remains analytic. In this paper, we employ a scheme with second-order
spatial accuracy for which finite difference and finite volume discretizations
are identical. However, since it is well known that finite difference is
computationally more efficient in multidimensions, we describe the numerical
method with finite difference representation. This makes it easy to extend the
scheme to higher orders if desired. The semi-discrete form of equations may be
integrated using any stable time integration schemes. In the numerical
examples shown in section 4, we always adopt the third order TVD Runge-Kutta
method of \cite{1988JCoPh..77..439S}.

\subsection{Discretization in Space}
\label{sec:discretization}

Let us consider a cartesian uniform mesh with sizes $\Delta x$, $\Delta y$,
$\Delta z$ in each direction. We define the fluid conservative variables
$\mathbf{U}$ (as point-value representations) at cell centers. In contrast,
three components of the magnetic field vector $B_x, B_y, B_z$ are defined at
staggered locations. Namely, each component of the magnetic field is defined
at face centers along the normal direction (say, $x$ direction for
$B_x$). Here by a face center we mean the center of the two-dimensional (2D)
plane that defines the interface between neighboring cells in a specific
direction. The same staggered discretization of the magnetic field is often
employed both in PIC and MHD simulation codes. The reason for this is that the
divergence-free condition for the magnetic field can be automatically
preserved up to machine accuracy when the induction equation is integrated in
time using the numerical flux (electric field) defined at edge centers (the
center of the line separating two neighboring faces). The staggering technique
applied to MHD is specifically referred to as the CT method
\citep{1988ApJ...332..659E}. In the next subsection, we discuss how the
magnetic field update accommodates consistently with an approximate Riemann
solver used to advance conservative variables defined at cell centers.

\subsection{HLL-UCT Scheme}
\label{sec:hll_uct}

Let us first consider a one-dimensional (1D) hyperbolic conservation law:
\begin{align}
 \frac{\partial \mathbf{u}}{\partial t} +
 \frac{\partial \mathbf{f}}{\partial x} = 0.
\end{align}
Temporal evolution of the solution vector $\mathbf{u}_i$ ($i$ indicates the
index for a cell) defined at cell centers may be described by the following
equation in the semi-discrete form:
\begin{align}
 \frac{d}{d t} \mathbf{u}_i + \frac{1}{\Delta x} \left(
 \hat{\mathbf{f}}_{i+1/2} - \hat{\mathbf{f}}_{i-1/2} \right) = 0.
\end{align}
The numerical flux $\hat{\mathbf{f}}_{i+1/2}$ is defined such that the above
difference equation gives an approximation to the spatial derivative with the
desired accuracy. Note that although this equation looks similar to finite
volume discretization, they are generally different when higher than second
order schemes are concerned. (In a finite volume scheme, $\mathbf{u}_i$ and
$\hat{\mathbf{f}}_{i+1/2}$ in the above equation should be replaced by the
cell average $\bar{\mathbf{u}}_i$ and point value $\mathbf{f}_{i+1/2}$,
respectively.) This illustrates only a 1D conservation law, but extension to
multidimensions is straightforward as far as the semi-discrete form combined
with finite difference discretization is employed. In contrast to this, the
situation becomes much more complicated when finite volume discretization is
used in multidimensions. In any case, since below we only consider a second
order scheme, these two approaches are identical. Nevertheless, the difference
must be kept in mind to be prepared for extension to higher orders.

The key question is how to evaluate the numerical flux
$\hat{\mathbf{f}}_{i+1/2}$. A typical strategy is to reconstruct the left and
right states of the solution vector at the cell interface as point-value
representations, which we denote by $\mathbf{u}_{i+1/2}^{\rmL}$ and
$\mathbf{u}_{i+1/2}^{\rmR}$, respectively. One may then solve the Riemann
problem at the cell interface either exactly or approximately to obtain the
numerical flux. In general, the exact Riemann solver is very expensive and
usually approximate Riemann solvers are adopted. Although a lot of Riemann
solvers have been proposed over the decades for the ideal MHD equations
\cite[e.g.,][]{1988JCoPh..75..400B,1998ApJS..116..119B,2005JCoPh.208..315M,2008ASPC..385..279M},
most of them cannot be applied to the QNTF model as they rely on the
eigenstructure of the basic equations. Furthermore, since the presence of
dispersive waves having characteristic temporal and spatial scales is an
intrinsic nature of the QNTF model, the solution to the Riemann problem is no
longer self-similar, making the situation much more complex.

To avoid complication inherent in the physical model, we adopt the HLL
approximate Riemann solver which does not require eigenmode decomposition.
The numerical flux in this approximation (as a point-value representation) is
given by
\begin{align}
 \mathbf{f} = \frac{\alphap \mathbf{f}^{\rmL} + \alpham \mathbf{f}^{\rmR}
 - \alphap \alpham (\mathbf{u}^{\rmR} - \mathbf{u}^{\rmL})} {\alphap + \alpham}
 \label{eq:hll}
\end{align}
where $\mathbf{f}^{\rm L,R} = \mathbf{f}(\mathbf{u}^{\rm L,R})$ and
$\alpha^{\pm}$ represent the maximum characteristic speeds (defined as
absolute values) in the positive and negative directions, respectively. In
general, $\mathbf{f}$ and $\hat{\mathbf{f}}$ are different and the correction
must be taken into account for higher than second order schemes
\citep[e.g.,][]{2003A&A...400..397D,2007A&A...473...11D}. However, in a second
order scheme used in this paper, taking $\mathbf{f} \approx \hat{\mathbf{f}}$
is sufficient. Note that we omit indices for $\alpha^{\pm}$, but they must
always be evaluated at the boundary where the Riemann problem is defined.

The HLL flux is obtained by assuming that the physical state is constant over
the Riemann fan. Thus, the only spectral information required for the solver
is the expansion velocities of the Riemann fan, which are estimated by the
maximum characteristic speeds $\alpha^{\pm}$. The scheme is also known as the
central-upwind scheme \citep{Kurganov:2001:SCS:587161.587347}, which is by
construction free from characteristic decomposition. Note that when the
symmetry over the Riemann fan $\alpha^{+} = \alpha^{-}$ is further assumed, it
reduces to the well-known LLF (local Lax-Friedrichs) flux. In this case, it is
also referred to as the central scheme \citep{2000JCoPh.160..241K}.

Now we consider application to the QNTF model with the CT-type discretization
described in the previous subsection. Although the CT scheme has been widely
used with many different Riemann solvers as well as reconstruction techniques,
it is not a trivial question how to couple the magnetic fields defined at face
centers and cell centers. If one tries to apply a 1D scheme to a
multidimensional problem via dimension by dimension (either with dimensionally
spit or unsplit) approach, the magnetic field must be defined at the same
location as other conservative variables. One may then obtain numerical fluxes
at face centers for each direction. On the other hand, the CT-type
discretization requires the numerical flux (i.e., electric field) defined at
edge centers. One immediately notices that since the electric fields at edge
centers are not available from a 1D Riemann solver used to solve the fluid
conservative variables, interpolation is needed to obtain the electric field
at edge centers. This leads to many different variants of methods that have
been proposed in the literature
\citep[e.g.,][]{1998ApJ...509..244R,1998JCoPh.142..331D,1999JCoPh.149..270B}.

The UCT framework gives a consistent way to calculate the numerical flux at
edge centers. It is actually designed to be consistent with an underlying
Riemann solver used to compute numerical fluxes for fluid conservative
variables. For simplicity, below we consider a 2D version of the scheme in the
$x{\rm -}y$ plane, but extension to 3D is trivial. The induction equation may
be written as
\begin{align}
 &\frac{d}{dt} B_{x;i+1/2,j} - \frac{c}{\Delta y} \left(
 \hat{E}_{z;i+1/2,j+1/2} - \hat{E}_{z;i+1/2,j-1/2} \right) = 0, \\
 &\frac{d}{dt} B_{y;i,j+1/2} + \frac{c}{\Delta x} \left(
 \hat{E}_{z;i+1/2,j+1/2} - \hat{E}_{z;i-1/2,j+1/2} \right) = 0.
\end{align}
This form clearly suggests that the numerical flux $\hat{E}_{z;i+1/2,j+1/2}$
must be defined in a genuinely multidimensional manner because it
simultaneously provides the flux for $B_x$ in $y$ direction and $B_y$ in $x$
direction \citep[see also,][]{2005JCoPh.205..509G}.

\cite{2004JCoPh.195...17L} proposed the following formula to calculate the
electric field at edge centers
\begin{align}
 E_{z} = \frac { \alphaxp \alphayp E_{z}^{\rmL_x \rmL_y} + \alphaxp \alphaym
 E_{z}^{\rmL_x \rmR_y} + \alphaxm \alphayp E_{z}^{\rmR_x \rmL_y} + \alphaxm
 \alphaym E_{z}^{\rmR_x \rmR_y}} {(\alphaxp + \alphaxm)(\alphayp + \alphaym)}
 + \frac{\alphaxp \alphaxm}{\alphaxp + \alphaxm} \frac{B_y^{\rmR_x} -
 B_y^{\rmL_x}}{c} - \frac{\alphayp \alphaym}{\alphayp + \alphaym}
 \frac{B_x^{\rmR_y} - B_x^{\rmL_y}}{c},
 \label{eq:ez_hll_uct1}
\end{align}
where the superscript indicates left and right states in $x$ and $y$
directions, respectively. (For instance, $\rmL_x$ represents the left state in
$x$ direction.) The maximum characteristic speeds in $x$ and $y$ directions
respectively are denoted by $\alpha_{x}^{\pm}$ and $\alpha_{y}^{\pm}$. Again,
$E_z \approx \hat{E}_z$ is satisfied to second order, but correction must be
taken into account for a higher order approximation. It is clear that the
above flux formula involves four states rather than two in the 1D HLL
flux. Therefore, this provides a flux of fully 2D in nature. Indeed, this
coincides with a 2D central-upwind (or HLL) flux formula for a Hamilton-Jacobi
equation given by \cite{Kurganov:2001:SCS:587161.587347}. This is actually to
be expected because the induction equation in 2D may be recognized as a
Hamilton-Jacobi equation in terms of the vector potential.

The crucial point in the above flux formula is that it automatically and
correctly reduces to the 1D HLL flux when homogeneity in one direction is
assumed. It is thus called the HLL-UCT scheme \citep{2004JCoPh.195...17L}.
The second and third terms play the role for the upwind property, but had been
ignored in earlier attempts to combine Riemann solvers with the CT
discretization considering only averaging of the electric field (i.e., the
first term). Notice that more elaborated multidimensional Riemann solvers have
been presented recently in the literature
\citep[e.g.,][]{2010JCoPh.229.1970B,2012JCoPh.231.7476B}. In this respect, the
HLL-UCT may be regarded as one of (and the simplest version of) the 2D Riemann
solvers.

Although the original work seems to be using implementation specific to the
ideal MHD Ohm's law \cite[see,][]{2007A&A...473...11D}, application of the UCT
scheme should not be restricted to a specific form of Ohm's law. Indeed, once
the electric field is determined at cell center (see section
\ref{sec:efield}), the numerical flux $\hat{E}_{z;i+1/2,j+1/2}$ may be
obtained by an appropriate reconstruction. However, the above form is not
necessarily convenient as it formally involves 2D interpolation of the
electric field to edge center points. In the next subsection, we introduce
slightly different definition of the numerical flux without loosing its
advantage, which can more easily be implemented using successive 1D
reconstructions.

\subsection{Reconstruction}
\label{sec:reconstruction}

One has to consider spatial reconstruction of physical quantities to calculate
the numerical fluxes. Throughout this paper, we use a piecewise linear
polynomial with the Monotonized Central (MC) slope limiter for non-oscillatory
reconstruction, which is therefore second order in space. On the other hand,
interpolation of the magnetic field defined at face centers (i.e., primary
data) to cell centers is performed as follows
\begin{align}
 &B_{x;i,j} = \frac{1}{2} \left( B_{x;i+1/2,j} + B_{x;i-1/2,j} \right) \\
 &B_{y;i,j} = \frac{1}{2} \left( B_{y;i,j+1/2} + B_{y;i,j-1/2} \right),
\end{align}
which is also correct to second order accuracy and does not involve a
nonlinear slope limiter. We note that there must be a correction term when one
employs the divergence-free reconstruction technique as proposed by
\cite{2001JCoPh.174..614B,2004ApJS..151..149B}. Nevertheless, we have not
implemented it yet at present.

We take $(\rho_i, \mathbf{v}_i, p_i, \rho_e, \mathbf{v}_e, p_e, \mathbf{E},
\mathbf{B})$ defined at cell centers as the variables for which the
reconstruction is performed. Recall that the density and velocity of ions and
electrons are not independent quantities. However, since the conversion from
ion to electron velocity involves calculation of $\rot \mathbf{B}$ (i.e.,
spatial derivatives), it is rather convenient to determine the electron
velocity first at cell centers which is then mapped to face centers via
reconstruction for calculation of the numerical flux. Although this
potentially introduces an inconsistency between ion and electron quantities
(which must be related by the constraints
Eqs.~(\ref{eq:current}-\ref{eq:charge})) reconstructed at face centers, we
have not encountered any difficulties associated with this. For the same
reason, we perform reconstruction for the electric field defined at cell
centers, for which one needs to solve the generalized Ohm's law (see section
\ref{sec:efield}).

Now consider calculation of the numerical flux $\mathbf{F}_x$ at each $x$-face
for the fluid conservative variables $\mathbf{U}$. This requires the left and
right states of fluid primitive variables $\rho_i, \mathbf{v}_i, p_i, \rho_e,
\mathbf{v}_e, p_e$ and the {\it transverse} components of electromagnetic
field $E_y, E_z, B_y, B_z$. Note that one can use the normal component of the
magnetic field $B_x$ already defined at this point without any reconstruction
(hence no ambiguity), whereas the electric field $E_x$ is not needed in the
flux calculation. One now obtains the numerical flux $\mathbf{F}_x$ using the
reconstructed left and right states and the HLL flux formula
Eq.~(\ref{eq:hll}). The same procedure is applied in the $y$ direction to
obtain $\mathbf{F}_y$.

In computing the numerical flux $\mathbf{F}_{x,y}$, one also calculates $E_z$
at each face as an appropriate average using available reconstructed left and
right states. Writing the reconstruction procedure to obtain the left and
right states symbolically as
\begin{align}
 f_{i+1/2}^{\rm L,R} \equiv {\mathcal R}_{i+1/2}^{\rm L,R} (f_i),
\end{align}
we use the following HLL average
\begin{align}
 &\langle E_{z;i,j} \rangle_{i+1/2} \equiv
 \frac {
 \alphaxp {\mathcal R}^{\rmL_x}_{i+1/2} \left( E_{z;i,j} \right) +
 \alphaxm {\mathcal R}^{\rmR_x}_{i+1/2} \left( E_{z;i,j} \right)
 }{\alphaxp + \alphaxm}
 \label{eq:hll_avg_x}
 \\
 &\langle E_{z;i,j} \rangle_{j+1/2} \equiv
 \frac {
 \alphayp {\mathcal R}^{\rmL_y}_{j+1/2} \left( E_{z;i,j} \right) +
 \alphaym {\mathcal R}^{\rmR_y}_{j+1/2} \left( E_{z;i,j} \right)
 }{\alphayp + \alphaym},
 \label{eq:hll_avg_y}
\end{align}
for the averaged electric field defined at $x$ and $y$ faces,
respectively. Here the angle bracket $\langle \rangle$ indicates the HLL
average along the direction specified by the subscript ($i+1/2, j+1/2$
respectively indicate averaging over $x$ and $y$ directions.)

These HLL-averaged electric fields are further reconstructed and averaged in
the other direction. Consequently, the electric field defined at edge center
$E_{z;i+1/2,j+1/2}$ is obtained as
\begin{align}
 E_{z;i+1/2,j+1/2} = &
 \frac{1}{2} \left\{
 \big \langle \langle E_{z;i,j} \rangle_{i+1/2} \big \rangle_{j+1/2} +
 \big \langle \langle E_{z;i,j} \rangle_{j+1/2} \big \rangle_{i+1/2}
 \right\}
 \nonumber \\
 &+ \frac{\alphaxp \alphaxm}{\alphaxp + \alphaxm}
 \frac{ B_{y;i+1/2,j+1/2}^{\rm R_x} - B_{y;i+1/2,j+1/2}^{\rm L_x} }{c}
 - \frac{\alphayp \alphaym}{\alphayp + \alphaym}
 \frac{ B_{x;i+1/2,j+1/2}^{\rm R_y} - B_{x;i+1/2,j+1/2}^{\rm L_y} }{c},
 \label{eq:hll_uct2}
\end{align}
where
\begin{align}
 B^{\rm L_y, R_y}_{x;i+1/2,j+1/2} \equiv {\mathcal R}^{\rm L_y, R_y}_{j+1/2}
 \left( B_{x;i+1/2,j} \right), \qquad
 B^{\rm L_x, R_x}_{y;i+1/2,j+1/2} \equiv {\mathcal R}^{\rm L_x, R_x}_{i+1/2}
 \left( B_{y;i,j+1/2} \right)
\end{align}
are results of 1D reconstruction of the magnetic field. Here the first term
represents the arithmetic mean of successive 1D reconstruction-averaging
procedures. For instance, the first term in the curly bracket of
Eq.~(\ref{eq:hll_uct2}) represents the averaging in the $x$ direction followed
by $y$ direction. It is readily be seen that this numerical flux reduces to
the original definition Eq.~(\ref{eq:ez_hll_uct1}) for the first-order
piecewise constant reconstruction with constant maximum characteristic
speeds. For a smooth region without discontinuities, one may also assume that
a direct 2D reconstruction and successive 1D reconstructions to the edge center
point will give the same result within the order of accuracy of
reconstruction. Although they do not necessarily coincide in general when a
nonlinear non-oscillatory reconstruction is used, it is rather important that
the above numerical flux retains the upwind property. It is again easy to
confirm that by taking the 1D limit this definition also reduces to the
correct 1D HLL flux. Therefore, we think it takes into account the essential
feature for the UCT scheme. Our numerical experiments performed with
Eq.~(\ref{eq:hll_uct2}) actually support this argument.

\subsection{Calculation of Electric Field}
\label{sec:efield}

The generalized Ohm's law used in this study is written in an implicit
form. Therefore, when discretized on a mesh, one must solve a matrix equation
to obtain the electric field. Notice that once the magnetic field and fluid
moment quantities are given, the equation is linear and thus solvable using
any matrix solvers. As we have already seen in section
\ref{sec:model_characteristics}, we can reasonably assume $\div \mathbf{E}
\approx 0$ for low frequency phenomena in a non-relativistic plasma. We thus
adopt the approximation
\begin{align}
 \nabla \times \nabla \times \mathbf{E} \approx - \nabla^2 \mathbf{E}
\end{align}
throughout in this paper.

We define all the three components of the electric field at cell
centers. Since we employ a second order scheme for solving the basic
equations, second order central difference discretization is sufficient for
approximating the spatial derivative in the generalized Ohm's law. Thus, the
left-hand side of the equation may be written as
\begin{align}
 \left( \Lambda \efd - c^2 \nabla^2 \efd \right)_{i,j,k} \approx &
 \left(\Lambda_{i,j,k} + 2(\epsilon_x + \epsilon_y + \epsilon_z)\right)
 \efd_{i,j,k} \nonumber \\ & - \epsilon_x \left( \efd_{i-1,j,k} +
 \efd_{i+1,j,k} \right) - \epsilon_y \left( \efd_{i,j-1,k} + \efd_{i,j+1,k}
 \right) - \epsilon_z \left( \efd_{i,j,k-1} + \efd_{i,j,k+1} \right)
\end{align}
where $\epsilon_x = c^2/\Delta x^2, \epsilon_y = c^2/\Delta y^2, \epsilon_z =
c^2/\Delta z^2$ and $\efd = E_x, E_y, E_z$ represents the three components of
the electric field vector. Likewise, the second order central difference is
used to evaluate spatial derivatives in the right-hand side, which gives the
source vector to the matrix equation. The discretization of the source term
may possibly cause even-odd decoupling, but so far we have not found any
problems.

The resulting matrix equation is iteratively solved using the symmetric
Gauss-Seidel method. Although it is known that the convergence of the method
is not so fast in general, it is sufficient for our purpose because the matrix
equation is diagonally dominant for the parameter regime of our interest. More
specifically, the ratio between the diagonal and non-diagonal coefficients is
proportional to $h^2/(c^2/\omega_{pe}^2)$ where $h$ represents the grid
size. Therefore, as far as the scale size large compared to the electron
inertial length is concerned, it is essentially a diagonal matrix and the
inversion is trivial. This is natural because the generalized Ohm's law is no
longer implicit in the absence of finite electron inertia. In most of
numerical examples shown in section 4, the grid size is comparable to or
larger than the electron inertial length, for which the matrix is relatively
easy to invert. The convergence is checked by monitoring the residual during
the iteration. It converges within a few iterations for a normalized tolerance
of $10^{-3}$ in most of the problems.

In cases where the electron inertial length is resolved by several grid
points, sometimes $30$ or more iterations are needed. If one is interested in
the electron scale physics, it is desirable to adopt a more advanced numerical
method such as multigrid. Nevertheless, our primary motivation is to introduce
a physical upper bound to the phase speed of whistler waves for better
numerical stability. For this reason, we have not yet tried to implement such
complicated methods.

\subsection{Characteristic Speed}
\label{sec:characteristic}

The only remaining task is to evaluate the maximum characteristic speed
required to determine the width of the Riemann fan. In principle, it may be
obtained by solving a linearized eigenvalue problem at the cell
interface. However, it will involve much more cumbersome algebra than for the
ideal MHD case. Furthermore, non-self-similarity of the problem makes such an
approach impractical. Therefore, we here adopt a simplified approach as
explained below.

We first evaluate the wave phase speed with respect to the plasma rest frame
by performing standard linear analysis of the basic equations for a
homogeneous plasma (see, \ref{sec:linear} for detail). After tedious
but straightforward algebra, one finds that a wave phase speed $X$ normalized
to the \Alfven speed $V_A = B/\sqrt{4 \pi(\rho_i + \rho_e)}$ may be obtained
by solving the 6th order polynomial equation
\begin{align}
 P X^6 - Q X^4 + R X^2 - S = 0
 \label{eq:lindisp}
\end{align}
with the coefficients defined as follows
\begin{align}
 & P = \left(1 + \varepsilon \kappa^2 \right)^2, \\
 & Q =
 1 + \beta + (1 + \kappa^2) \cos^2 \theta +
 \varepsilon \kappa^2 \left(1 - \cos^2 \theta + 2 \beta \right) +
 \varepsilon^2 \kappa^2 \left( \cos^2 \theta + \beta \kappa^2 \right),
 \\
 & R =
 \left( 1 + (2 + \kappa^2 ) \beta + \varepsilon^2 \kappa^2 \beta \right)
 \cos^2 \theta, \\
 & S = \beta \cos^4 \theta.
\end{align}
Here $\varepsilon \equiv m_e/m_i$, $\beta \equiv 4\pi \gamma (p_i + p_e)/B^2$
and $\theta$ represents the wave propagation angle with respect to the ambient
magnetic field. Note that the wavenumber $\kappa$ is normalized to
$\Omega_{ci}/V_A$ which differs from the inverse ion inertia length by a
factor $1/\sqrt{1 + \varepsilon}$. It is easy to confirm that the
characteristic equation reduces to the ideal MHD dispersion relation by taking
the long wavelength limit $\kappa \rightarrow 0$. A finite wavenumber $\kappa
\neq 0$ thus describes the two-fluid effect. The dispersion relation of
Hall-MHD is obtained in the limit $\varepsilon \rightarrow 0$ (i.e.,
negligible electron inertia).

In the following, we always evaluate the phase speed at the Nyquist
wavenumber. This choice is motivated from the conjecture that the Riemann fan
should be bounded by the maximum possible characteristic speeds, which
typically correspond to the wave at the Nyquist (i.e., largest) wavenumber. We
note this does not hold when the grid size is much smaller than the electron
inertial length. Nevertheless, we have confirmed that a different choice of
the wavenumber does not introduce appreciable difference in our numerical
results.

A root with the maximum absolute value of the above equation gives the maximum
wave phase speed in the plasma rest frame. A reasonable estimate for the
expansion speed of the Riemann fan may be obtained by transforming it back to
the laboratory frame. For this, one has to find the ``bulk velocity'' with
which the bulk fluid moves with respect to the laboratory frame. This is,
however, not a trivial problem in the present model. Namely, in situations
where a finite current flows, the ion and electron flow velocities are
different. The effect is more and more pronounced where the ion and electron
dynamics is decoupled and the Hall current plays the role. This ambiguity
arises perhaps because our perturbation analysis is performed for a
homogeneous plasma where the current is assumed to be zero. Again, our
strategy is to estimate an upper limit of the characteristic speeds. For this,
we adopt
\begin{align}
 \alpha^{\pm} = \max \left\{X V_A \pm v_i, X V_A \pm v_e, 0\right\}
\end{align}
as the maximum characteristic speeds, where $v_i$ and $v_e$ are ion and
electron bulk velocities normal to the interface.

The characteristic equation is cubic with respect to $X^2$, which can easily
be solved by an iterative method. Note that since we can use the solution
obtained at the previous time step as the initial guess (which is likely to be
very close to the solution), iterative methods would be superior to analytical
calculation. We use the standard Newton method for root finding. It is easy to
confirm from elementary calculus that the desired root should be found in the
range $X^2 > \left(Q + \sqrt{Q^2 - 3PR}\right)/(3P)$.

Our approach is obviously not exact but gives a measure for the upper bound to
the expansion speed of the Riemann fan. This implies that numerical
dissipation is slightly increased with this approach. Nevertheless, the
ambiguity in the definition of the Riemann fan appears only when the
dispersive effect becomes important. In this case, a shock is no longer a pure
discontinuity but has a structure of finite width accompanied by dispersive
waves. For this reason, we think that the increased numerical dissipation is
not a critical issue in this model.

\subsection{Summary of Numerical Procedure}
\label{sec:summary}

Here we summarize the procedure used in our simulation code. We initialize the
fluid primitive variables and the magnetic field at edge centers, such that
$\div {\mathbf B} = 0$ is satisfied. The cell-center magnetic field is
calculated by interpolation, and the primitive variables are converted to the
conservative variables at cell centers.

At the beginning of the Runge-Kutta integration, the maximum characteristic
speed is calculated at cell centers, and is assumed to be constant during the
one Runge-Kutta time step. Appropriate mapping of the characteristic speed
from cell centers to face/edge centers is performed. We take the maximum of
the two neighboring characteristic speeds in the normal direction to obtain
the estimate at each face. Then, their simple arithmetic average in the
transverse direction is taken to evaluate the values at edge center.

With this preparation, the following six steps are performed for each substep
of the Runge-Kutta time integration:
\begin{enumerate}
 \item The electric field is calculated at cell centers by solving the
       generalized Ohm's law Eq.~(\ref{eq:ohm}).
 \item Reconstruction of fluid primitive variables and transverse
       electromagnetic field components defined at cell center is performed to
       obtain the left and right states at face centers for each direction.
 \item Numerical fluxes for the fluid conservative variables are calculated
       using the HLL flux formula Eq.~(\ref{eq:hll}) at each face. At the same
       time, the HLL average of the transverse electric field components are
       calculated according to the definition
       Eqs.(\ref{eq:hll_avg_x}-\ref{eq:hll_avg_y}) and are stored on a working
       array.
 \item Reconstruction of the electromagnetic field (i.e, the primary magnetic
       field and HLL-averaged electric field) defined at face center is
       performed. The numerical fluxes for the induction equation are
       calculated using the formula Eq.~(\ref{eq:hll_uct2}).
 \item The fluid conservative variables at cell center as well as magnetic
       field at face center are updated in a conservative manner using the
       numerical fluxes obtained above.
 \item The updated magnetic field at face centers is interpolated to cell
       center. The fluid primitive variables are then recovered at cell center
       from the updated conservative variables.
\end{enumerate}
This completes the description of the numerical procedure. Since our numerical
algorithm follows earlier studies using the HLL-UCT with finite difference
framework \citep{2003A&A...400..397D,2007A&A...473...11D} except for
implementation specific to MHD, it is relatively easy to extend the code to
higher orders. Implementation of higher order schemes is left for future
studies.

\section{Numerical Results}
\label{sec:results}

In this section, we discuss numerical results for several benchmark problems
obtained with our new simulation code for the QNTF equations. Since we think
it may replace the role of Hall-MHD and/or EMTF equations in plasma
simulations, it is fair to employ benchmark problems that have been well
tested with these models. We have found that, however, test problems for
Hall-MHD models have not necessarily been well established yet; most of
previous studies tested their codes against simple whistler wave propagation
and the well-known GEM (Geospace Environment Modeling) magnetic reconnection
challenge problems. In particular, problems involving shocks and/or other
discontinuities in multidimensions have not been tested in the published
literature. This may be related to the fact that many Hall-MHD codes implement
numerical dissipation for short wavelength modes in an ad-hoc manner for
numerical stability.

We thus adopt benchmark problems that have been tested with the EMTF and MHD
models. Since the QNTF model correctly reproduces the MHD result if the grid
size is appropriately chosen, benchmark problems for MHD can be naturally
extended to the regime where two-fluid effect becomes apparent. Concerning
numerical results obtained with the EMTF model, we expect the difference
should be negligible whenever the charge neutrality assumption is adequate.
Therefore, for low-frequency ($\omega \ll \omega_{pe}$) problems, it is
reasonable to compare our numerical solutions with the published EMTF results.

Since we know the maximum characteristic speed, the time step is chosen such
that it always satisfies the CFL (Courant-Friedrichs-Lewy) stability criterion
in the entire system. In the numerical examples shown in this paper, we always
use a time step with a CFL number less than $0.5$. More specifically, when the
CFL number becomes greater than 0.5 (smaller than 0.25), the time step is
halved (doubled).

In the numerical examples shown below, we use a polytropic index of $\gamma =
5/3$, resistivity $\eta = 0$, and temperature ratio $\tau = 1$ unless
otherwise stated. Since by construction our scheme automatically satisfies the
solenoidal condition for the magnetic field, the $\div \mathbf{B}$ errors are
not shown explicitly. We have confirmed that the error is indeed always at a
level of machine accuracy. Although our code is fully 3D, only 1D and 2D
simulation results are discussed below because they are sufficient to
demonstrate the capability of the code with modest computational time.

\subsection{Brio-Wu Shock Tube}
\begin{figure}[t]
 \begin{center}
  \includegraphics[scale=0.75]{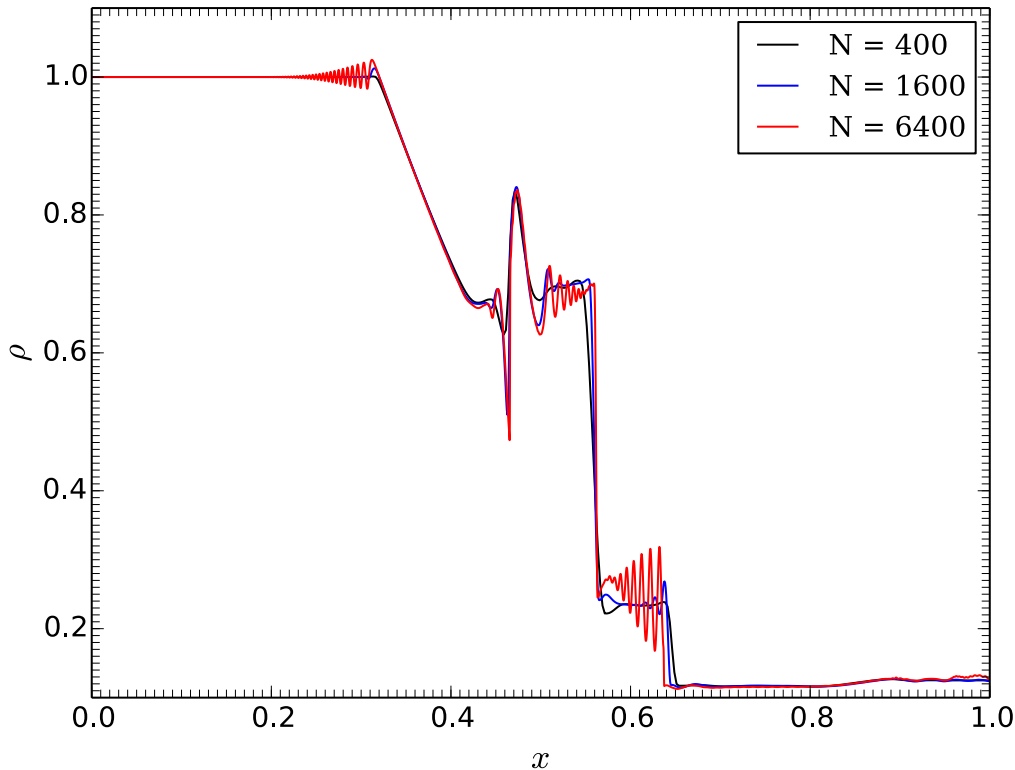}

  \caption{Total mass density for the Brio-Wu shock tube problem at $t =
  0.1$. Simulation results with three different resolutions are shown with
  black ($N = 400$), blue ($N = 1600$), and red ($N = 6400$) lines,
  respectively.}

  \label{fig:bw-density}
 \end{center}
\end{figure}

\begin{figure}[t]
 \begin{center}
  \includegraphics[scale=0.75]{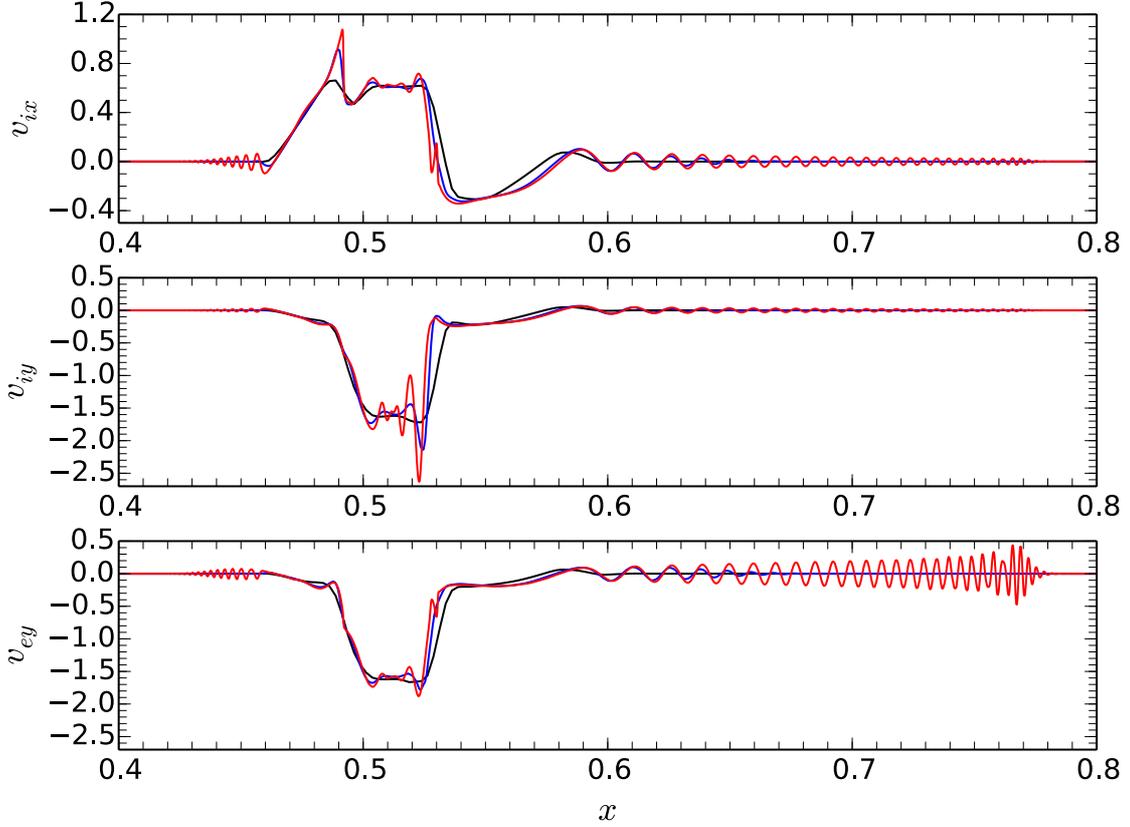}

  \caption{Closeup view of ion and electron velocities for the Brio-Wu shock
  tube problem at $t = 0.02$. The $x$ (top) and $y$ (middle) components of ion
  velocity and $y$ component of electron velocity (bottom) are shown. The
  meaning of the color is the same as the previous figure.}

  \label{fig:bw-velocity}
 \end{center}
\end{figure}

The first test problem is an extension of 1D Riemann problem proposed by
\cite{1988JCoPh..75..400B} that is one of the standard test problems for the
ideal MHD. It has also been tested by the EMTF model, which clearly
demonstrates the appearance of dispersive features
\citep[e.g.,][]{2006JCoPh.219..418H,Kumar2012}.

At the initial condition, the simulation box of unit length $0 \leq x \leq 1$
was divided into the left and right states at the center $x = 0.5$. Their
physical quantities were specified respectively as follows
\begin{align}
 \begin{pmatrix}
  \rho \\ p \\ B_x/\sqrt{4\pi} \\ B_y/\sqrt{4\pi}
 \end{pmatrix}_{\rm left}
 =
 \begin{pmatrix}
  1.0 \\ 1.0 \\ 0.75 \\ 1.0 \\
 \end{pmatrix},
 \qquad
 \begin{pmatrix}
  \rho \\ p \\ B_x/\sqrt{4\pi} \\ B_y/\sqrt{4\pi}
 \end{pmatrix}_{\rm right}
 =
 \begin{pmatrix}
  0.1 \\ 0.125 \\ 0.75 \\ -1.0
 \end{pmatrix},
\end{align}
where $\rho \equiv \rho_i + \rho_e$ and $p \equiv p_i + p_e$ are the total
mass density and total pressure, respectively.  Other quantities were zero
everywhere at the initial condition. This setup is identical to the original
Brio-Wu shock tube problem for the ideal MHD. However, the parameter $q_i/m_i$
can be chosen arbitrarily to control the ion inertial length $\lambda_i
\propto (q_i/m_i)^{-1}$. Clearly, the ideal MHD corresponds to the limit
$q_i/m_i \rightarrow \infty \, (\lambda_i \rightarrow 0)$. Here we chose
$q_i/m_i$ such that $\lambda_i = 10^{-3} / \sqrt{\rho}$. In this case, the
problem is close to MHD but two-fluid nature will appear when sufficiently
small grid size is used to resolve $\lambda_i$. For this problem, $\gamma = 2$
was used instead of $5/3$. The ion to electron mass ratio was chosen to be
$m_i/m_e = 100$.

Fig.~\ref{fig:bw-density} shows snapshots of the total mass density $\rho$ at
$t = 0.1$ obtained with three different resolutions $N = 400, 1600, 6400$
(i.e., $\Delta x = 2.5 \times 10^{-3}, 6.25 \times 10^{-4}, 1.5625 \times
10^{-4}$). At the lowest resolution (black), the grid size was comparable to
or larger than the ion inertial length and dispersive nature was not clearly
visible. Consequently, the numerical solution was almost the same as the
MHD. In contrast, a clear dispersive wave train structure appeared behind the
slow shock at around $x \sim 0.65$ in the medium resolution run
(blue). Similar structures were also found at the leading edge of fast mode
rarefaction propagating to the left ($x \sim 0.3$), as well as around a
compound wave ($x \sim 0.5$). These features became further pronounced in the
solution with the highest resolution. It is interesting that such dispersive
waves are clearly visible even with such a small ion inertial length. We found
that the propagation speed of the slow shock appears to be slower than the MHD
result. This seems to be consistent with the result presented by
\cite{2006JCoPh.219..418H}. This may be understood to be due to a finite
Poynting flux carried by the dispersive waves behind the shock.

It is noted that at this time ($t = 0.1$) whistler waves emitted to the right
had already reached to the right-hand boundary, which is, however, barely
visible in the mass density profile. The propagation of whistler waves can
more clearly be identified in the velocity profiles at earlier times. Closeup
view of the ion and electron velocities at $t = 0.02$ shown in
Fig.~\ref{fig:bw-velocity} displays the propagation of whistler waves in both
positive and negative directions again for three different resolutions. Note
that the normal component of ion and electron velocities are identical in 1D
because $\left(\nabla \times \mathbf{B}\right)_{x} = 0$.

We see that as increasing the spatial resolution, the leading edge of whistler
waves became more and more extended. In particular for the whistlers
propagating to the right, the wavelength of the transverse electron velocity
oscillations decreased in the positive $x$ direction. This is clearly due to
dispersive nature of whistler waves. Namely, since the whistler waves have
dispersion relation $\omega \propto k^2$ for $|\Omega_{ci}| \ll \omega \ll
|\Omega_{ce}|$, the group velocity becomes higher for shorter wavelength
modes. Note that the maximum wave propagation velocity (both group and phase
velocity) was limited by finite electron inertia effect in the highest
resolution run because the grid size was comparable to the electron inertial
length for this simulation ($m_i/m_e = 100$). If the spatial resolution is
insufficient to resolve these small-scale features, numerical dissipation
automatically damps out these waves.

Here, we have actually confirmed that our numerical simulation code
automatically reduces essentially to ideal MHD when the grid size is larger
than the ion inertial length. Since high frequency waves such as
electromagnetic and Langmuir waves are absent, numerical stability only
requires the CFL condition with respect to MHD characteristic speeds. We think
this is a crucial advantage over the EMTF model.

\subsection{Circularly Polarized Wave}
\begin{figure}[t]
 \begin{center}
  \includegraphics[scale=0.75]{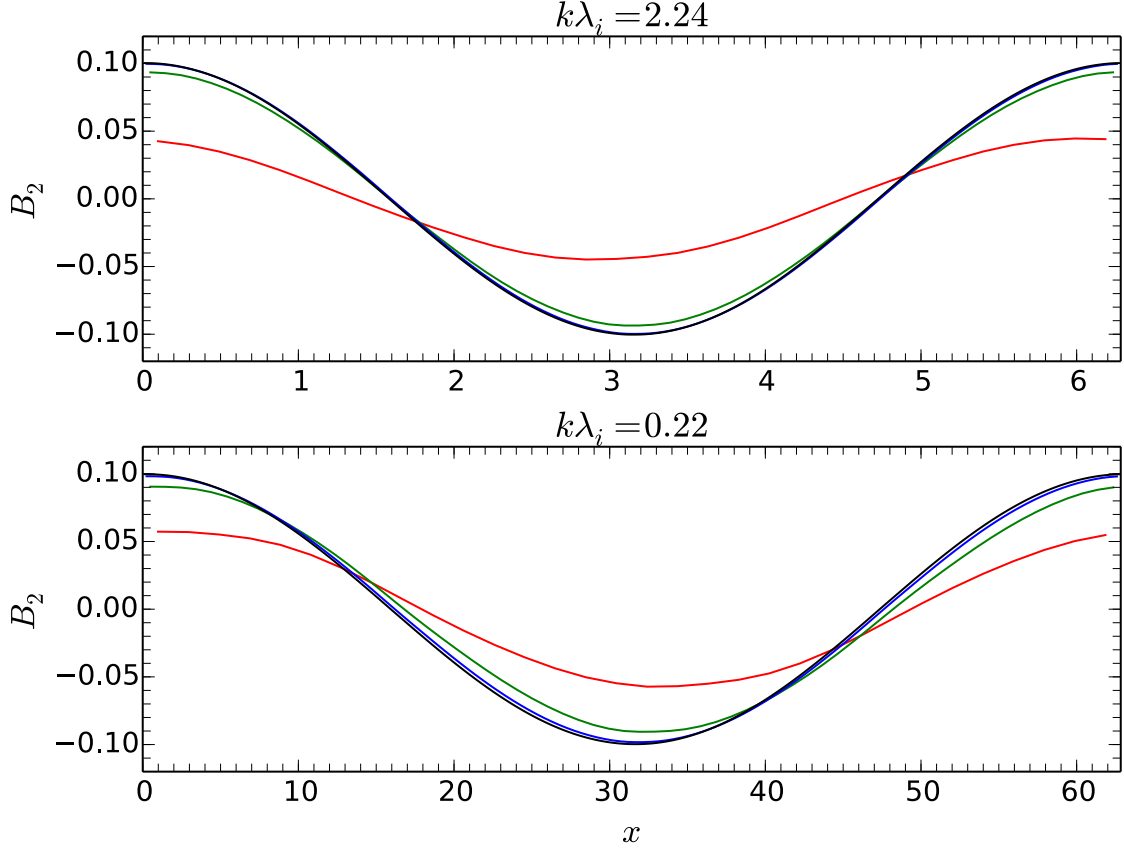}

  \caption{Profiles of the in-plane wave magnetic field component obtained by
  the circularly polarized wave problem. The highest frequency case $k
  \lambda_i \approx 2.24$ (top), and lowest frequency case $k \lambda_i
  \approx 0.22$ (bottom) are shown respectively. Results for four different
  resolutions are shown with red ($N = 16$), green ($N = 32$), blue ($N =
  64$), and black ($N = 128$) lines, respectively. Note that the blue and
  black lines almost coincide in this figure. Likewise, the exact solutions
  are indistinguishable from the highest resolution results.}

  \label{fig:cpw-profile}
 \end{center}
\end{figure}

\begin{figure}[t]
 \begin{center}
  \includegraphics[scale=0.75]{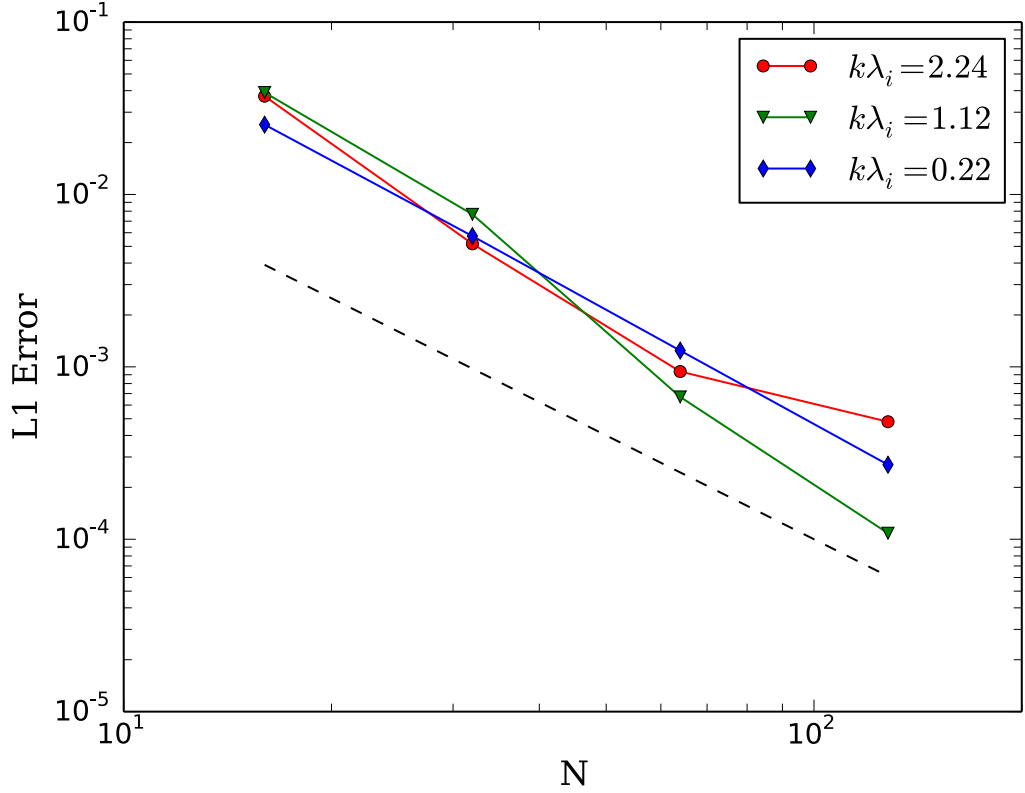}

  \caption{Convergence of L1 error for the in-plane wave magnetic field
  component as a function of the number of grid $N$ for the circularly
  polarized wave problem. Convergence for three different wavenumber cases is
  shown with different colors: red for $k \lambda_i \approx 2.24$, green for
  $k \lambda_i = 1.12$, and blue for $k \lambda_i \approx 0.22$. The black
  dashed line indicates the theoretical second order convergence for
  reference.}

  \label{fig:cpw-error}
 \end{center}
\end{figure}

It has been well known that a finite amplitude circularly polarized \Alfven
wave propagating along the ambient magnetic field ($\mathbf{B}_0$) is an exact
solution to the ideal MHD equations
\citep[e.g.,][]{1978ApJ...219..700G}. Therefore, it is commonly used as a
benchmark problem to test the accuracy of a numerical scheme.

Generalization of the MHD exact solution to the QNTF model is
possible. Actually, the dispersion relation of a finite amplitude circularly
polarized electromagnetic wave is identical to the corresponding linear
dispersion relation
\begin{align}
 1
 + \left( \frac{\omega_{pi}}{k c} \right)^2 \frac{\omega}{\omega + \Omega_{ci}}
 + \left( \frac{\omega_{pe}}{k c} \right)^2 \frac{\omega}{\omega + \Omega_{ce}}
 = 0,
\end{align}
where $\Omega_{cs} = q_s B_0 / m_s c$ is the cyclotron frequency. The positive
(negative) $\omega$ in the above dispersion relation indicates right-handed
(left-handed) polarization. Adopting a coordinate system with orthogonal unit
vectors $\mathbf{e}_i \, (i=1,2,3)$ such that $\mathbf{e}_1$ is parallel to
the background field, $\mathbf{e}_2$ is a vector perpendicular to
$\mathbf{e}_1$, and $\mathbf{e}_3 = \mathbf{e}_1 \times \mathbf{e}_2$, we can
write the eigenvector corresponding to the solution of the dispersion relation
as follows
\begin{align}
 & B_2 = \xi B_0 \cos(k x - \omega t), \quad
 B_3 = \xi B_0 \sin(k x - \omega t), \\
 & v_{s,2} = V_{s} \cos(k x - \omega t), \quad
 v_{s,3} = V_{s} \sin(k x - \omega t),
\end{align}
where
\begin{align}
 V_{s} = - \xi \frac{\Omega_{cs}}{\omega + \Omega_{cs}} \frac{\omega}{k}.
\end{align}
The amplitude of the wave $\xi$ normalized to the background magnetic field
$B_0$ is a free parameter that can be chosen arbitrarily. Nevertheless, since
the finite amplitude wave may suffer a parametric instability
\citep[e.g.,][]{1978ApJ...219..700G,1986JGR....91.4171T}, we chose a small value
$\xi = 0.1$ to suppress possible growth of the instability during the
simulation.

We set up a 2D simulation box with the computational domain $0 \leq x \leq
2L$, $0 \leq y \leq L$ with the periodic boundary condition in both
directions. The simulation box was resolved by $2N \times N$ grid points
(i.e., $\Delta x = \Delta y$). The background magnetic field was inclined with
respect to the $x$ axis by an angle $\theta = \tan^{-1} (2)$. The wavenumber
of a finite amplitude circularly polarized wave was chosen as $(k_x, k_y) =
(\pi/L, 2 \pi/L)$ (i.e., mode = 1 in each direction). The initial condition
was set up by rotating the exact solution by an angle $\theta$ such that
$\mathbf{e}_1$ is parallel to the background magnetic field and $\mathbf{e}_2$
is in the simulation plane. Since the wave propagates oblique to the grid,
this provides a benchmark problem to test the accuracy of handling smooth
profiles in multidimensions. In this simulation, time and length were
normalized to the ion cyclotron frequency $\Omega_{ci}$ and ion inertial
length $\lambda_i = c/\omega_{pi}$, respectively. The ambient magnetic field
strength was chosen to be $B_0 = \sqrt{4 \pi}$.

By appropriately choosing $L/\lambda_i$, one may test the code performance for
different parameter regimes. We here discuss three different physical
wavenumber cases, $k \lambda_i \approx 2.24, 1.12, 0.22$ (corresponding to
$L/\lambda_i = \pi, 2 \pi, 10 \pi$, respectively). The wave frequencies were
respectively determined as $\omega/\Omega_{ci} \approx 5.72, 1.89, 0.25$ by
solving the dispersion relation for right-handed polarization ($\omega >
0$). The lowest frequency case is essentially an \Alfven wave in the MHD
regime, whereas higher frequency cases correspond to dispersive whistler
waves. In all cases, a plasma beta of $\beta = 0.1$, and an ion to electron
mass ratio of $m_i/m_e = 256$ were used.

Fig.~\ref{fig:cpw-profile} shows the profiles of the in-plane component of
wave magnetic field $B_2 = (-2 B_x + B_y)/\sqrt{5}$ as functions of $x$ (i.e.,
cut through $y = 0$) after five wave periods. Results with the highest and
lowest frequency runs are shown in top and bottom panels, respectively. The
amplitude is normalized to the background field $B_0$. The simulations were
performed with four different resolutions $N = 16, 32, 64, 128$ for each
case. We see that in both cases the numerical solutions converged to the
analytic ones as increasing the resolution. Note that the initial conditions
(or exact solutions) are indistinguishable from the highest resolution results
and are not shown.

The error convergence is shown in Fig.~\ref{fig:cpw-error} for three different
wavenumbers. The L1 errors shown in this plot are those of the $B_2$ component
evaluated as
\begin{align}
 L1(B_2) = \frac{1}{2N^2 B_0} \sum_{i,j}
 \left|B_{2;i,j} (t=5T) - B_{2;i,j} (t=0)\right|,
\end{align}
where $T$ denotes a wave period. This confirms that the code achieved roughly
second order convergence (which is shown by the dashed line for
reference). However, the error behavior became irregular for the higher
frequency cases at high resolution. We have performed simulations with other
wavenumbers and/or mass ratios, and confirmed that this irregular convergence
appears when the grid size becomes comparable to or smaller than the electron
inertial length.  This may thus partly be related to the simplified estimate
of the maximum characteristic speeds that is not strictly correct at this
scale. In any case, since our primary focus is not on this small scale, we
have not attempted to resolve the issue.

\subsection{Orszag-Tang Vortex}
\begin{figure}[t]
 \begin{center}
  \includegraphics[scale=0.50]{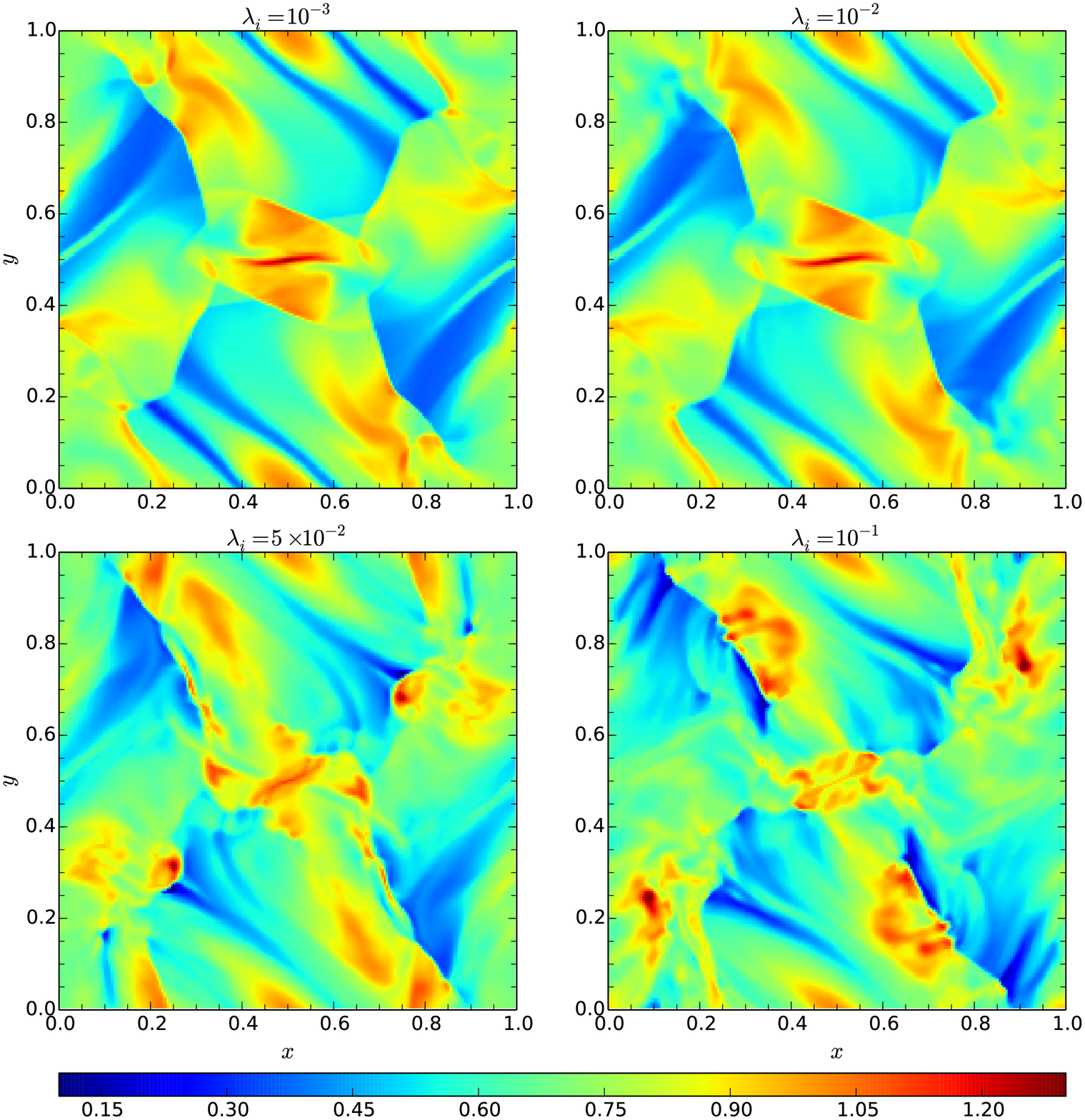}

  \caption{Temperature distribution for the Orszag-Tang vortex problem at $t =
  0.5$. The solutions with different initial ion inertial lengths are shown;
  $\lambda_i = 10^{-3}$ (top left), $10^{-2}$ (top right), $5 \times 10^{-2}$
  (bottom left), $10^{-1}$ bottom right.}

  \label{fig:ot-temperature}
 \end{center}
\end{figure}

\begin{figure}[t]
 \begin{center}
  \includegraphics[scale=0.50]{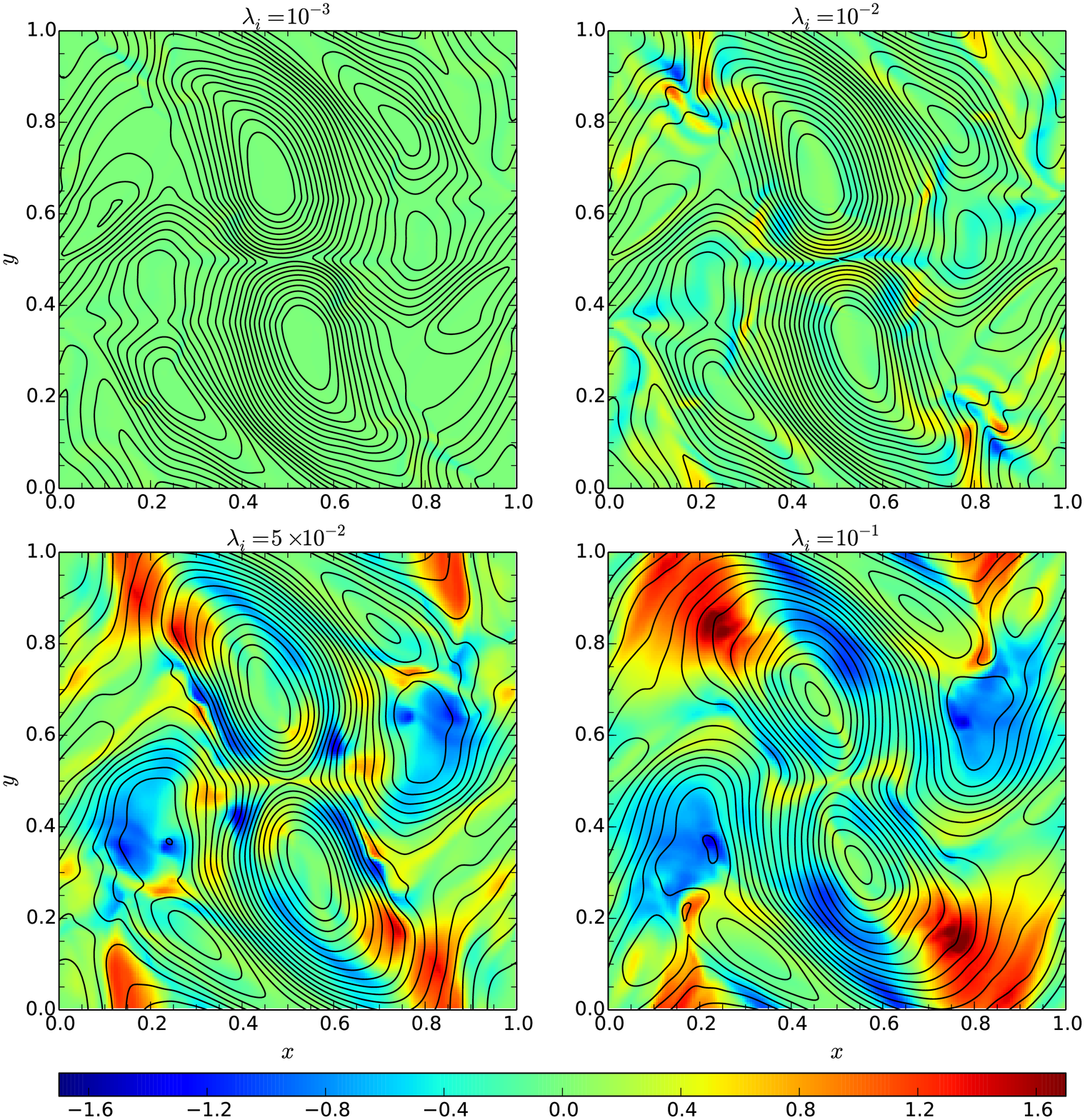}

  \caption{Out-of-plane magnetic field component $B_z$ and the vector
  potential $A_z$ for the Orszag-Tang vortex problem at $t = 0.5$. The color
  indicates $B_z$, while black lines are contours of $A_z$ (magnetic field
  lines in the $x{\rm -}y$ plane), respectively. The format is the same as the
  previous figure.}

  \label{fig:ot-bfield}
 \end{center}
\end{figure}

The Orszag-Tang vortex problem has been one of the most standard benchmark
problems for multidimensional MHD codes. It starts with a smooth initial
profile, but the solution soon becomes complex involving many
discontinuities. Therefore, it has been used to test the capability of
handling multidimensional discontinuities which may introduce non-negligible
magnetic monopoles in the numerical solution. On the other hand, to the
authors knowledge, numerical results obtained with the Hall-MHD and/or EMTF
models have not been published in the literature. Although there were
numerical studies with similar but different initial conditions
\citep[e.g.,][]{1995LNP...462..399M,2012PhRvL.109s1101P}, they are not useful
for comparison with the present results. Therefore, we generalize the original
problem to the parameter regime where the Hall term starts to play a role, but
without changing the initial condition so that comparison with published
results becomes easier.

The computational domain was a unit square $0 \leq x \leq 1$, $0 \leq y \leq
1$ with the periodic boundary condition in both directions. The initial
condition was given by the smooth profile
\begin{align}
 &
 v_{x} = - \sin \left(2 \pi y\right), \quad
 v_{y} =   \sin \left(2 \pi x\right), \\
 &
 B_{x} = - B_0 \sin \left(2 \pi y\right), \quad
 B_{y} =   B_0 \sin \left(4 \pi x\right),
\end{align}
where $B_0 = \sqrt{4 \pi}$. The density and pressure were constant; $\rho =
\gamma^2, p = \gamma$, and out-of-plane velocity and magnetic field components
were zero $v_{z} = B_{z} = 0$. This definition is identical to the standard
MHD test problem (except for normalization). Again, one needs to specify the
ion inertial length (or $q_i/m_i$) relative to the box size.

Numerical results with four different initial ion inertial lengths $\lambda_i
= 10^{-3}, 10^{-2}, 5 \times 10^{-2}, 10^{-1}$ are shown in
Figs.~\ref{fig:ot-temperature} and \ref{fig:ot-bfield}. We used $200 \times
200$ grid points and $m_i/m_e = 100$ in all the simulation
runs. Fig.~\ref{fig:ot-temperature} displays snapshots of the temperature
distribution at $t = 0.5$. (Here, the temperature is defined with the total
pressure and mass density as $p/\rho$.) In Fig.~\ref{fig:ot-bfield}, the color
indicates the out-of-plane component of the magnetic field normalized to
$B_0$, whereas the black solid lines are contours of the vector potential
$A_z$ (i.e., magnetic field lines in the plane) at the same instant of
time. We confirmed that the $\nabla \cdot \mathbf{B}$ error in the numerical
solutions was on the order of round-off error, which is consistent with the
design of the numerical scheme.

Since the grid size was $\Delta x = 5 \times 10^{-3}$, the run with the
smallest ion inertial length $\lambda_i = 10^{-3}$ did not resolve the ion
scale. It was thus essentially in the MHD regime and the results may be
compared with published results for ideal MHD. We see that our simulation
well captured discontinuities and other small scale features. This is not
surprising because in this case our scheme becomes almost the same as the
second order scheme described in \cite{2004JCoPh.195...17L} except for
subtleties. Although the overall structures were similar in all the cases,
dispersive features clearly appeared in the numerical solutions as increasing
the ion inertial length. The fact that the Hall term played the role in the
numerical solutions can most clearly be visible in Fig.~\ref{fig:ot-bfield}
because the out-of-plane magnetic field disappears in the ideal MHD
limit. Even in the case with $\lambda_i = 10^{-2}$, in which the temperature
distribution looks almost unchanged from the MHD result, amplitude of $B_z$
became substantial (at around $x \sim 1.5, y \sim 0.9$ and its symmetric
point). For larger ion inertial lengths, the solution became much more complex
even in this early stage of evolution.

The numerical results clearly indicate that it is possible to implement the
two-fluid effect in a shock-capturing code without loosing its advantages.  A
shock-capturing code gives a non-oscillatory solution when encountered by
discontinuities by automatically introducing required dissipation. On the
other hand, dispersive character must be retained at ion and electron inertial
scales, which tends to produce an oscillatory solution. Our numerical
simulation code successfully implements these two contradictory features at
the same time without suffering from numerical instability.

\subsection{Magnetic Reconnection}

\begin{figure}[t]
 \begin{center}
  \includegraphics[scale=0.75]{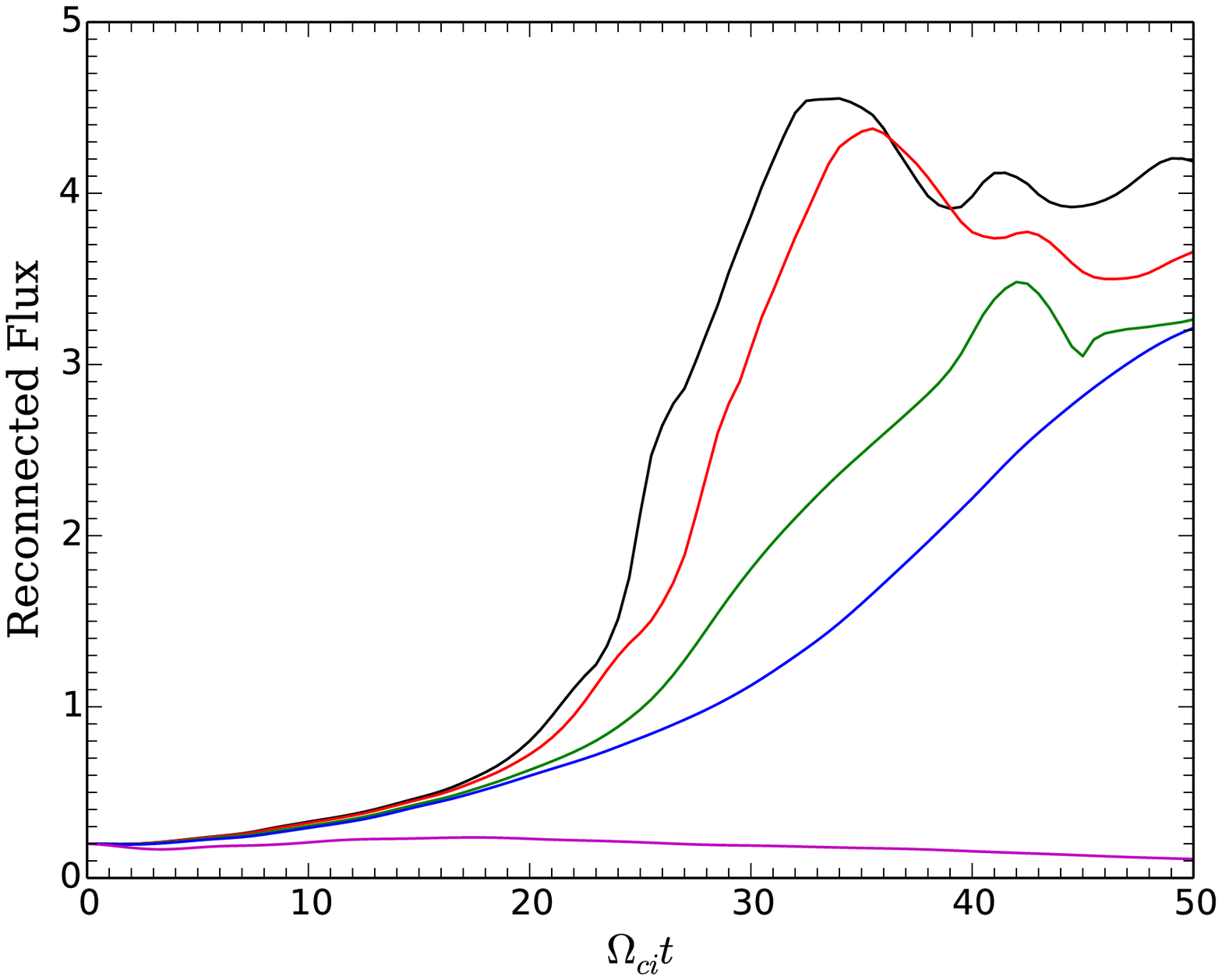}

  \caption{Time evolution of reconnected magnetic flux for the magnetic
  reconnection problem. Five runs with different normalized resistivities are
  shown in different colors. The black, red, green, blue, magenta lines
  correspond to $\eta^{\star} = 0, 10^{-6}, 5 \times 10^{-6}, 10^{-5},
  10^{-4}$, respectively.}

  \label{fig:mrx-flux}
 \end{center}
\end{figure}

\begin{figure}[t]
 \begin{center}
  \includegraphics[scale=0.50]{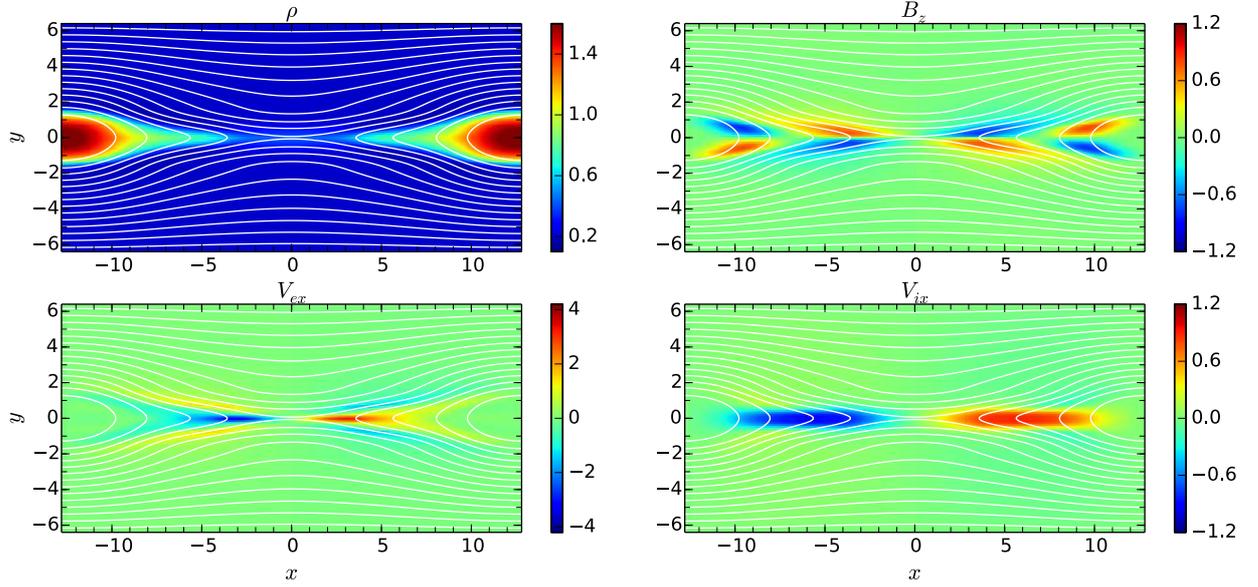}

  \caption{Snapshots for the magnetic reconnection problem at $\Omega_{ci} t =
  21.5$ for the run with $\eta^{\star} = 0$. Shown are the mass density (top
  left), out-of-plane magnetic field component (top right), $x$ component of
  electron velocity (bottom left), and ion velocity (bottom right). The white
  lines in each panel shows the magnetic field lines (contour lines for the
  vector potential $A_z$). At this time, the reconnected magnetic flux reached
  $\psi(t) = 1.0$.}

  \label{fig:mrx1-t1}
 \end{center}
\end{figure}

\begin{figure}[t]
 \begin{center}
  \includegraphics[scale=0.50]{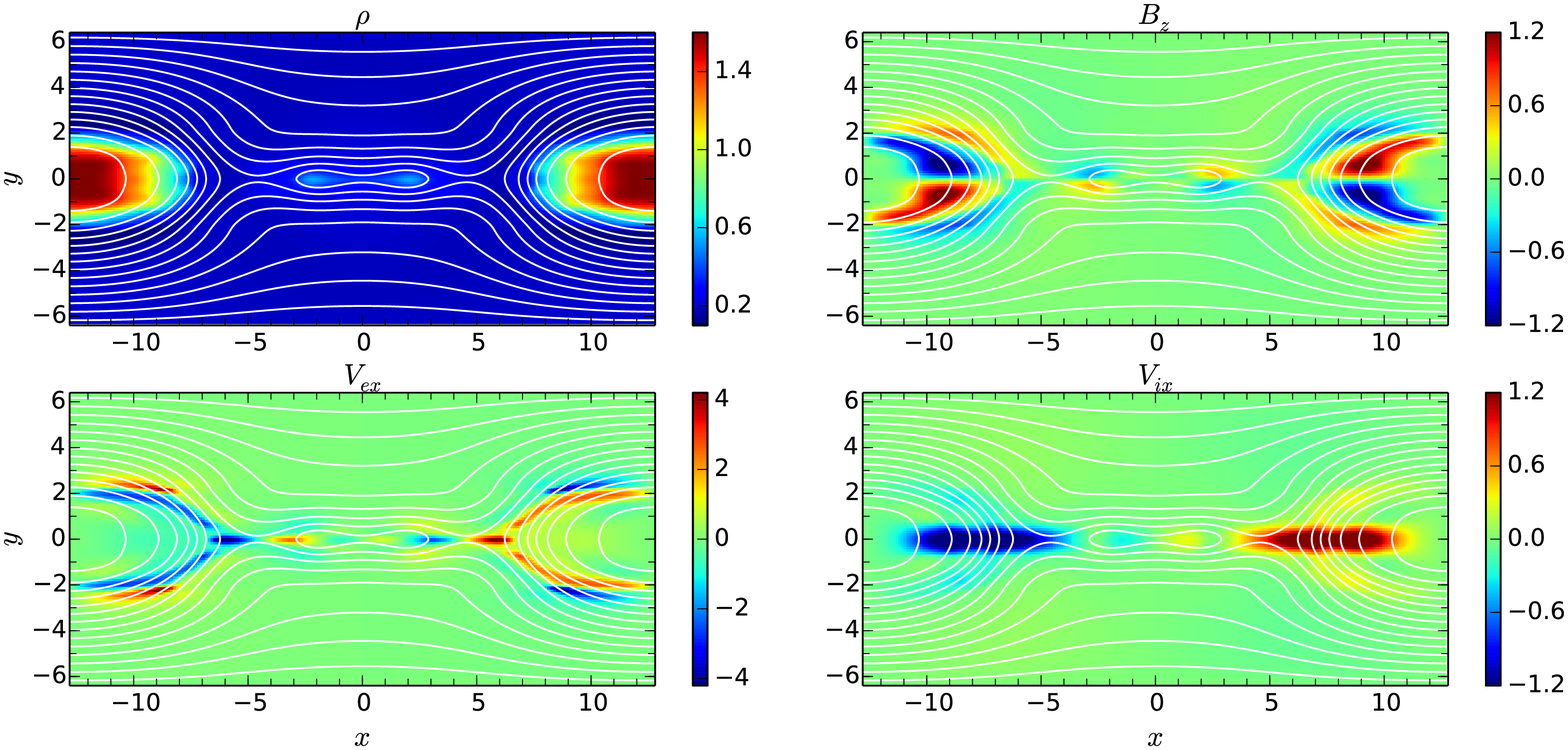}

  \caption{Snapshots for the magnetic reconnection problem at $\Omega_{ci} t =
  25.0$ (corresponds to the time $\psi(t) = 2.0$) for the run with
  $\eta^{\star} = 0$. The format is the same as the previous figure.}

  \label{fig:mrx1-t3}
 \end{center}
\end{figure}

\begin{figure}[t]
 \begin{center}
  \includegraphics[scale=0.50]{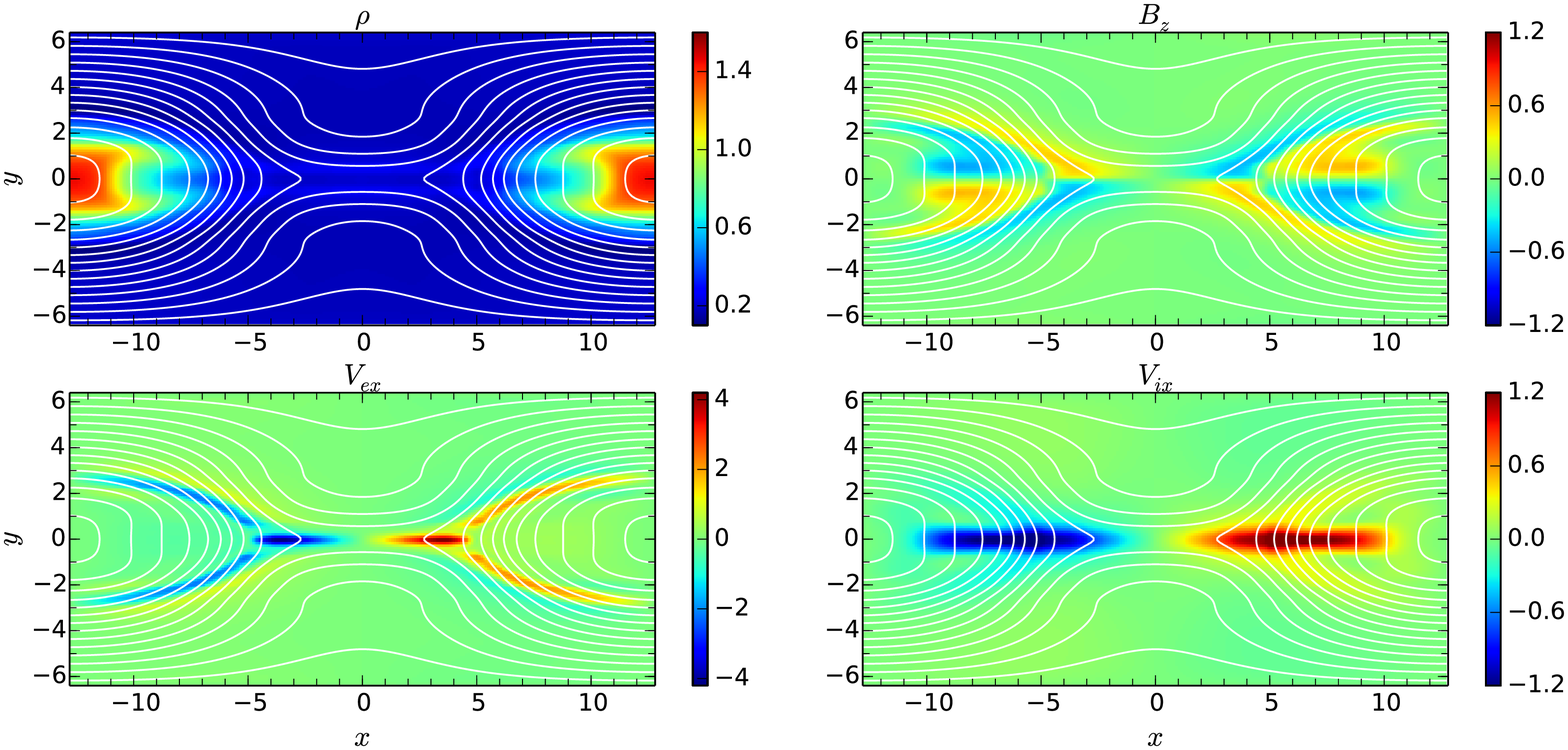}

  \caption{Snapshots for the magnetic reconnection problem at $\Omega_{ci} t =
  31.5$ (corresponds to the time $\psi(t) = 2.0$) for the run with
  $\eta^{\star} = 5 \times 10^{-6}$. The format is the same as the previous
  figure.}

  \label{fig:mrx3-t3}
 \end{center}
\end{figure}

Our final test problem is the GEM magnetic reconnection challenge problem that
has been well tested with many different simulation models
\citep{2001JGR...106.3715B}. Here we use this problem to test the effect of
resistivity. The initial condition was given by the Harris equilibrium which
is identical to the one described in \cite{2001JGR...106.3715B}. The magnetic
field and number density profiles were given by
\begin{align}
 B_x(y) = B_0 \tanh \left( y/d \right)
\end{align}
and
\begin{align}
 n(y)   = n_0 \sech^2 \left( y/d \right) + n_{bg},
\end{align}
respectively. Here, the width of the current sheet is represented by $d$. The
ion and electron temperatures were determined by the pressure balance
condition $n_0 (T_i + T_e) = B_0^2/8\pi$. We here used a temperature ratio of
$\tau = 5$ to be consistent with the original problem. The ion and electron
drift velocities in the $z$ direction must satisfy the relation $\tau =
-v_{i,z}/v_{e,z}$ for the initial condition to be a Vlasov-Maxwell
equilibrium.  The drift velocities were thus initialized as
\begin{align}
 &v_{i,z} = -
 \frac{c B_0}{4 \pi e} \frac{\tau}{1 + \tau}
 \frac{n_0 \sech^2 (y/d)}{n_0 \sech^2 \left( y/d \right) + n_{bg}}, \\
 &v_{e,z} = +
 \frac{c B_0}{4 \pi e} \frac{1}{1 + \tau}
 \frac{n_0 \sech^2 (y/d)}{n_0 \sech^2 \left( y/d \right) + n_{bg}}
\end{align}
for ions and electrons, respectively. Other quantities were initialized with
zero.

The normalization of time and space was such that the ion cyclotron frequency
$\Omega_{ci} = e B_0/m_i c = 1$ and the ion inertial length $c/\omega_{pi} = c
\sqrt{m_i/4\pi n_0 e^2} = 1$. Accordingly, the velocity was normalized to the
ion \Alfven speed $V_{A,i} = B_0/\sqrt{4 \pi n_0 m_i}$. Note that in the
normalized unit, $B_0/\sqrt{4 \pi} = n_0 = m_i = 1$. We used a rectangular
simulation domain $-L \leq x \leq +L$ and $-L/2 \leq y \leq +L/2$ with $L =
12.8$. The current sheet thickness was taken to be $d = 0.5$. The simulations
were performed with $256 \times 128$ grid points with a mass ratio of $m_i/m_e
= 25$. The grid size $\Delta x = \Delta y = 0.1$ was thus roughly comparable
to the electron inertial length for the initial current sheet density. The
periodic boundary condition was used in $x$ direction, while the conducting
wall boundary was used in $y$ direction.

To initiate magnetic reconnection, the initial magnetic field was perturbed
with the out-of-plane component of the vector potential given by
\begin{align}
 \Phi = \Phi_0 \frac{\pi B_0}{L}
 \cos \left( \frac{\pi x}{L} \right) \cos \left( \frac{\pi y}{L} \right)
\end{align}
where $\Phi_0 = 0.1$. With this initial condition, the X-point was located
initially at the origin from which the reconnection process started to evolve.

Fig.~\ref{fig:mrx-flux} shows time evolution of reconnected magnetic flux for
simulations with five different normalized resistivities: $\eta^{\star} = 0,
10^{-6}, 5 \times 10^{-6}, 10^{-5}, 10^{-4}$. Recall that since the
resistivity is normalized with respect to the local plasma frequency, the
diffusivity depends on the local density even with a spatially constant
resistivity. The reconnected flux was computed by
\begin{align}
 \psi(t) = \frac{1}{2 B_0} \int_{-L}^{+L} |B_y(x, y=0, t)| dx.
\end{align}
In all cases except for $\eta^{\star} = 10^{-4}$, the reconnection rates
estimated from the slope of reconnected flux exceeded $\sim 0.1$ in the
nonlinear phase, consistent with published results.

Fig.~\ref{fig:mrx1-t1} shows snapshots at $\Omega_{ci} t = 21.5$ for
$\eta^{\star} = 0$, at which the normalized reconnected flux reached $\psi(t)
= 1.0$. Fast outflows originating from the X-point are clearly seen for both
ion and electron velocities. The ions and electrons were accelerated to the
\Alfven speeds defined for each species ($V_{A,i}$ for ions and $V_{A,e} =
\sqrt{m_i/m_e} V_{A,i}$ for electrons). The out-of-plane magnetic field $B_z$
was generated in association with the decoupling between ion and electron
dynamics. Similar characteristics were also found in other runs at the time
such that $\psi(t) = 1.0$, although the magnitude decreased with increasing
the resistivity. These features are consistent with previous numerical
studies.

There was a quantitative difference in the time evolution of reconnected flux
in between $\eta^{\star} = 10^{-6}$ and $5 \times 10^{-6}$ after $\Omega_{ci}
t \gtrsim 25$. We found that the further increase in the reconnection rate in
smaller resistivity runs was associated with the formation of secondary
magnetic islands. Figs.~\ref{fig:mrx1-t3} and \ref{fig:mrx3-t3} compare runs
with $\eta^{\star} = 0$ and $5 \times 10^{-6}$ at the time when the same
amount of flux was reconnected $\psi(t) = 2.0$ ($\Omega_{ci} t = 25.0$ and
$31.5$ respectively). Two newly formed X-points in the elongated thin current
sheet can be seen in Fig.~\ref{fig:mrx1-t3}, whereas the dynamics was still
dominated by a single X-point in Fig.~\ref{fig:mrx3-t3}. Formation of
secondary magnetic islands was not observed until the end of simulations
$\Omega_{ci} t = 50$ for $\eta^{\star} \geq 5 \times 10^{-6}$. This difference
may be understood from the viewpoint of the stability of the thin (i.e.,
electron-scale) current sheet generated in the nonlinear phase of reconnection
process. In higher resistivity runs, the length of the thin current sheet
characterized by the fast electron outflow became shorter because of the
diffusion during the propagation. This is probably the reason for prohibiting
the excitation of a secondary tearing mode. The electron-scale current sheet
may also become unstable against another instability in fully 3D simulations
\citep{1994PhRvL..73.1251D}. Nevertheless, one must bare in mind that electron
kinetic physics, which is not included in the present model, will play a role
in the dynamics of such a thin current sheet.

\section{Conclusions}
\label{sec:conclusions}

In the present paper, we have proposed the QNTF model for collisionless
plasmas. The basic equations are the fluid equations for ions and electrons
and the Maxwell equations without displacement current. The absence of the
displacement current indicates the charge neutrality, which is a reasonable
assumption for low frequency phenomena. Rewriting the basic equations, it is
shown that the system consists of the conservation laws for five scalar
variables (total mass, momentum, and energy), and the induction equation for
the magnetic field, supplemented by the generalized Ohm's law. The system
fully takes into account finite electron inertia effect, which introduces the
upper bound to the phase speed of whistler waves. On the other hand, it
reduces to the ideal MHD in the long wavelength limit. No high frequency waves
(such as Langmuir or electromagnetic waves) exist in the system, which would
otherwise impose a severe restriction on the simulation time step even in the
long wavelength limit.

We have developed a 3D numerical simulation code for solving the QNTF
equations. The code employs the HLL approximate Riemann solver combined with
the UCT method for the induction equation. With this approach, we avoid
complicated characteristic decomposition with keeping the advantage of a
shock-capturing scheme. The UCT scheme guarantees the divergence-free
condition for the magnetic field up to machine accuracy, without loosing
upwind property. Therefore, the code is able to capture complicated
multidimensional sharp discontinuities without appreciable numerical
oscillation. At the same time, it successfully describes dispersive waves
arising from the two-fluid effect without numerical problem. The upper bound
of whistler wave phase speed appropriately introduced by finite electron
inertia helps to overcome the well-known numerical stability issue in dealing
with the short wavelength mode.

In the numerical examples, we have confirmed that the numerical solutions
reduce to the ideal MHD results when the ion inertial length is small
enough. In this case, numerical stability requires only the CFL condition with
respect to the MHD characteristic speed because of the absence of high
frequency waves. This is in clear contrast to the EMTF equations in which high
frequency waves must be resolved by the simulation time step. Since these
waves will not play a major role in non-relativistic plasmas, the low
frequency approximation is adequate. Consequently, we believe that the QNTF
model offers a better alternative to the Hall-MHD and/or EMTF models.

\section*{Acknowledgments}
The author is indebted to T.~Minoshima, K.~Hirabayashi, and T.~Miyoshi for
helpful discussion. This work was supported by JSPS Grant-in-Aid for Young
Scientists (B) 25800101.


%% trick for figure numbering
\let\thefigureSAVED\thefigure
\appendix
\let\thefigure\thefigureSAVED

\section{Derivation of Generalized Ohm's Law}
\label{sec:ohmslaw}

We here introduce a collision term in the right-hand side of the equation of
motion for the two fluids to take into account finite resistivity $\eta$. It
is intended to be rather phenomenological (or anomalous) such as due to
kinetic wave-particle interactions. The collision term is defined as
\begin{align}
 \mathbf{R}_i = - \mathbf{R}_e =
 - \frac{\eta}{4 \pi} \frac{\omega_{pe}^2 + \omega_{pi}^2}
 {\dfrac{q_i}{m_i} - \dfrac{q_e}{m_e}}
 \left(
 \frac{q_i}{m_i} \rho_i \mathbf{v}_i + \frac{q_e}{m_e} \rho_e \mathbf{v}_e
 \right) =
 - \frac{\eta}{4 \pi} \frac{\omega_{p}^2}
 {\dfrac{q_i}{m_i} - \dfrac{q_e}{m_e}}
 \mathbf{J}
\end{align}
where $\mathbf{R}_i$ and $\mathbf{R}_e$ are for ion and electron fluids
respectively. By adding it into the equation of motion, one may take into
account effective friction between the species. It is easy to understand from
the symmetry that the definition does not violate the momentum conservation
law.

The generalized Ohm's law Eq.~(\ref{eq:ohm}) used in this paper may be
obtained by taking a weighted sum between the two equations with the weight
factor $q_s/m_s$. This yields
\begin{align}
 \frac{\partial}{\partial t} {\mathbf J} +
 \nabla \cdot \left[ \sum_{s} \frac{q_s}{m_s}
 \left( \rho_s {\mathbf v}_s {\mathbf v}_s + p_s {\mathbf I} \right) \right] =
 \sum_{s} \left[
 \rho_s \frac{q_s^2}{m_s^2} {\mathbf E} +
 \rho_s \frac{q_s^2}{m_s^2} \frac{{\mathbf v}_s}{c} \times {\mathbf B}
 \right]
 - \frac{\eta}{4 \pi} \omega_{p}^2 \mathbf{J}.
\end{align}
Using Faraday's law, the first term may be rewritten as follows
\begin{align}
 \frac{\partial}{\partial t} {\mathbf J}
 &= \frac{c}{4\pi} \nabla \times \frac{\partial}{\partial t} {\mathbf B}
 \nonumber \\
 &=-\frac{c^2}{4\pi} \nabla \times \nabla \times {\mathbf E}.
\end{align}
Now rearranging the equation, we arrive at
\begin{align}
 \left( \sum_{s} \omega_{ps}^2 \right) {\mathbf E} +
 c^2 \nabla \times \nabla \times {\mathbf E} = -
 \left( \sum_{s} \omega_{ps}^2 \frac{{\mathbf v}_{s}}{c} \right)
 \times {\mathbf B} +
 \nabla \cdot
 \left( \sum_{s} \frac{4 \pi q_s}{m_s}
 \left( \rho_s {\mathbf v}_{s} {\mathbf v}_{s} + p_s {\mathbf I} \right)
 \right) +
 \eta \omega_{p}^2 \mathbf{J},
\end{align}
which is identical to Eq.~(\ref{eq:ohm}). It should be noted that
contributions from both ion and electron fluids are fully taken into account
in this form of Ohm's law, so that it always provides the correct equation
regardless of the ion-to-electron mass ratio.

\section{Linear Dispersion Analysis}
\label{sec:linear}

Here we outline the linear dispersion analysis for the QNTF model. We consider
a homogeneous electron-proton plasma without resistivity, i.e., $q_i = -q_e =
e$ and $\eta = 0$. The ion fluid equations, the generalized Ohm's law, and the
induction equation are used. Hereafter, we drop the subscript for the particle
species unless necessary, and the fluid quantities without subscript must be
read as the ion quantities. In addition, we use the definition for the thermal
velocity $V_i^2 = \gamma k_B T_i/m_i, V_e^2 = \gamma k_B T_e/m_e$ (where $k_B$
is the Boltzmann constant) and the following notation: $\varepsilon =
m_e/m_i$, $\tau = T_i/T_e$, $\mu = \varepsilon (1 - \tau \varepsilon)/(1 +
\varepsilon)$. Note that in the derivation below, the assumption of constant
$\tau$ is not necessary.

It is easy to obtain the following equation from the linearized ion fluid
equations
\begin{align}
 \left(
 \omega^2 - V_{i}^2 \mathbf{k} \mathbf{k}
 \right) \cdot \delta \mathbf{v} =
 i \omega \frac{e}{m_i}
 \left(
 \delta \mathbf{E} + \frac{\delta \mathbf{v}}{c} \times \mathbf{B}_0
 \right),
 \label{eq:lin_ion}
\end{align}
whereas for the generalized Ohm's law, we have
\begin{align}
 \left( \omega_{p}^2 - c^2 \mathbf{k} \times \mathbf{k} \times \right)
 \delta \mathbf{E} = -
 \frac{\delta \mathbf{\Gamma}}{c} \times \mathbf{B}_0
 + i \mathbf{k} \cdot \delta \mathbf{\Pi},
\end{align}
with
\begin{align}
 & \delta \mathbf{\Gamma} =
 \omega_{p}^2
 \left\{
 \delta \mathbf{v} -
 \frac{m_i}{e} \frac{c}{4 \pi \rho_0 (1 + \varepsilon)}
 i \mathbf{k} \times \delta \mathbf{B}
 \right\}
 \\
 & \delta \mathbf{\Pi} = -\omega_{p}^2 \frac{m_i}{e}
 \mu V_{e}^2
 \frac{\mathbf{k} \cdot \delta \mathbf{v}}{\omega}.
\end{align}

Rearranging the above equation using $\delta \mathbf{B} = c \mathbf{k}/\omega
\times \delta \mathbf{E}$, we obtain
\begin{align}
 i \omega \frac{e}{m_i}
 \left(
 \delta \mathbf{E} + \frac{\delta \mathbf{v}}{c} \times \mathbf{B}_0
 \right) =
 \mu V_{e}^2 \mathbf{k} \mathbf{k} \cdot \delta \mathbf{v} +
 i \frac{e}{m_i} \lambda^2
 \left[
 \varepsilon \omega \mathbf{k} \times \mathbf{k} \times \delta \mathbf{E} -
 i \mathbf{\Omega}_{ci} \times \mathbf{k} \times \mathbf{k}
 \times \delta \mathbf{E}
 \right],
 \label{eq:lin_ohm}
\end{align}
where we have defined $\lambda^2 = V_A^2/\Omega_{ci}^2 = c^2/\omega_{pi}^2 (1
+ \varepsilon)$ and $\mathbf{\Omega}_{ci} = \Omega_{ci} \mathbf{B}_0/B_0$,
respectively.

Now our task is to eliminate $\delta \mathbf{v}$ from Eqs.~(\ref{eq:lin_ion})
and (\ref{eq:lin_ohm}), to get a dispersion matrix $\mathbf{D}$ satisfying
\begin{align}
 \mathbf{D} \cdot \delta \mathbf{E} = \mathbf{0},
\end{align}
from which the dispersion relation is naturally obtained as $|\mathbf{D}| =
0$. For this purpose, we introduce the following matrix notation:
\begin{align}
 \mathbf{M} & \equiv \omega^2 \mathbf{I} - V_S^2 \mathbf{k} \mathbf{k}
 \\
 k^2 \mathbf{N} \cdot \delta \mathbf{E} &\equiv
 i \mathbf{\Omega}_{ci} \times \mathbf{k} \times \mathbf{k}
 \times \delta \mathbf{E}
 - \varepsilon \omega
 \mathbf{k} \times \mathbf{k} \times \delta \mathbf{E}
 \\
 \mathbf{W} \cdot \delta \mathbf{v} &\equiv
 \mathbf{\Omega}_{ci} \times \delta \mathbf{v},
\end{align}
where the sound speed is defined by
\begin{align}
 V_S^2 \equiv
 V_{i}^2 + \mu V_{e}^2 =
 \frac{\gamma k_B (T_i + T_e)}{m_i + m_e}.
\end{align}
Solving the equations with respect to the ion velocity, we have
\begin{align}
 \delta \mathbf{v} = - i \frac{e}{m_i} k^2 \lambda^2
 \mathbf{M}^{-1} \cdot \mathbf{N} \cdot \delta \mathbf{E}.
\end{align}
Substituting this into the generalized Ohm's law Eq.~(\ref{eq:lin_ohm}), we
finally obtain
\begin{align}
 \mathbf{D} =
 \mathbf{I} + k^2 \lambda^2
 \left[
 i \mathbf{W} \cdot \mathbf{M}^{-1} \cdot \mathbf{N} +
 \mu \frac{V_{e}^2}{\omega} \mathbf{k} \mathbf{k}
 \cdot \mathbf{M}^{-1} \cdot \mathbf{N} +
 \frac{1}{\omega} \mathbf{N}
 \right]
\end{align}
as the dispersion matrix.

Explicit form of the normalized dispersion matrix is given as follows:
\begin{align}
 \mathbf{D} =
 \frac{1}{X^2 (X^2 - \beta)}
 \left[
 \mathbf{D}_{MHD} + \mathbf{D}_{i} + \mathbf{D}_{e}
 \right],
\end{align}
where
\begin{align}
 \mathbf{D}_{MHD} &=
 \begin{pmatrix}
  (X^2 - \beta^2)(X^2 - \cos^2 \theta) &
  0 &
  (X^2 - \beta^2) \cos \theta \sin \theta
  \\
  0 &
  X^4 - (\beta^2 + 1) X^2 + \beta^2 \cos^2 \theta &
  0 &
  \\
  0 &
  0 &
  X^2 (X^2 - \beta^2) &
 \end{pmatrix},
 \\
 \mathbf{D}_{i} &= i \kappa X
 \begin{pmatrix}
  0 &
  X^2 - \beta^2 + \mu \beta_e^2 \sin^2 \theta &
  0 &
  \\
  -(X^2 - \beta^2) \cos^2 \theta &
  0 &
  (X^2 - \beta^2) \cos \theta \sin \theta
  \\
  0 &
  \mu \beta_e^2 \cos \theta \sin \theta &
  0 &
 \end{pmatrix},
 \\
 \mathbf{D}_{e} &= i \varepsilon \kappa X (X^2 - \beta^2)
 \begin{pmatrix}
  - i \kappa X \cos^2 \theta &
  -1 &
  i \kappa X \cos \theta \sin \theta &
  \\
  \cos^2 \theta &
  - i \kappa X &
  - \cos \theta \sin \theta &
  \\
  i \kappa X \cos \theta \sin \theta &
  0 &
  -i \kappa X \sin^2 \theta &
 \end{pmatrix}.
\end{align}
In the above expression, $X = \omega/k V_A$ is the normalized phase speed,
$\kappa = k \lambda$ is the normalized wavenumber, $\theta$ is the wave
propagation angle with respect to the ambient magnetic field. We also defined
$\beta = V_S^2/V_A^2$ and $\beta_e = V_e^2/V_A^2$, respectively.

One may easily understand that $\mathbf{D}_{MHD}$ (which remains finite for
$\kappa \rightarrow 0$) corresponds to the MHD limit. On the other hand,
$\mathbf{D}_{i}$ (independent of $\varepsilon$) and $\mathbf{D}_{e}$ represent
respectively the ion and electron inertia effects. Taking the determinant, and
arranging it into the polynomial form, we obtain the dispersion relation given
in Eq.~(\ref{eq:lindisp}). Most of cumbersome calculation presented in the
derivation has been performed with a computer algebra system package SymPy
\citep{SymPy}.

\begin{figure}[t]
 \begin{center}
  \includegraphics[scale=0.50]{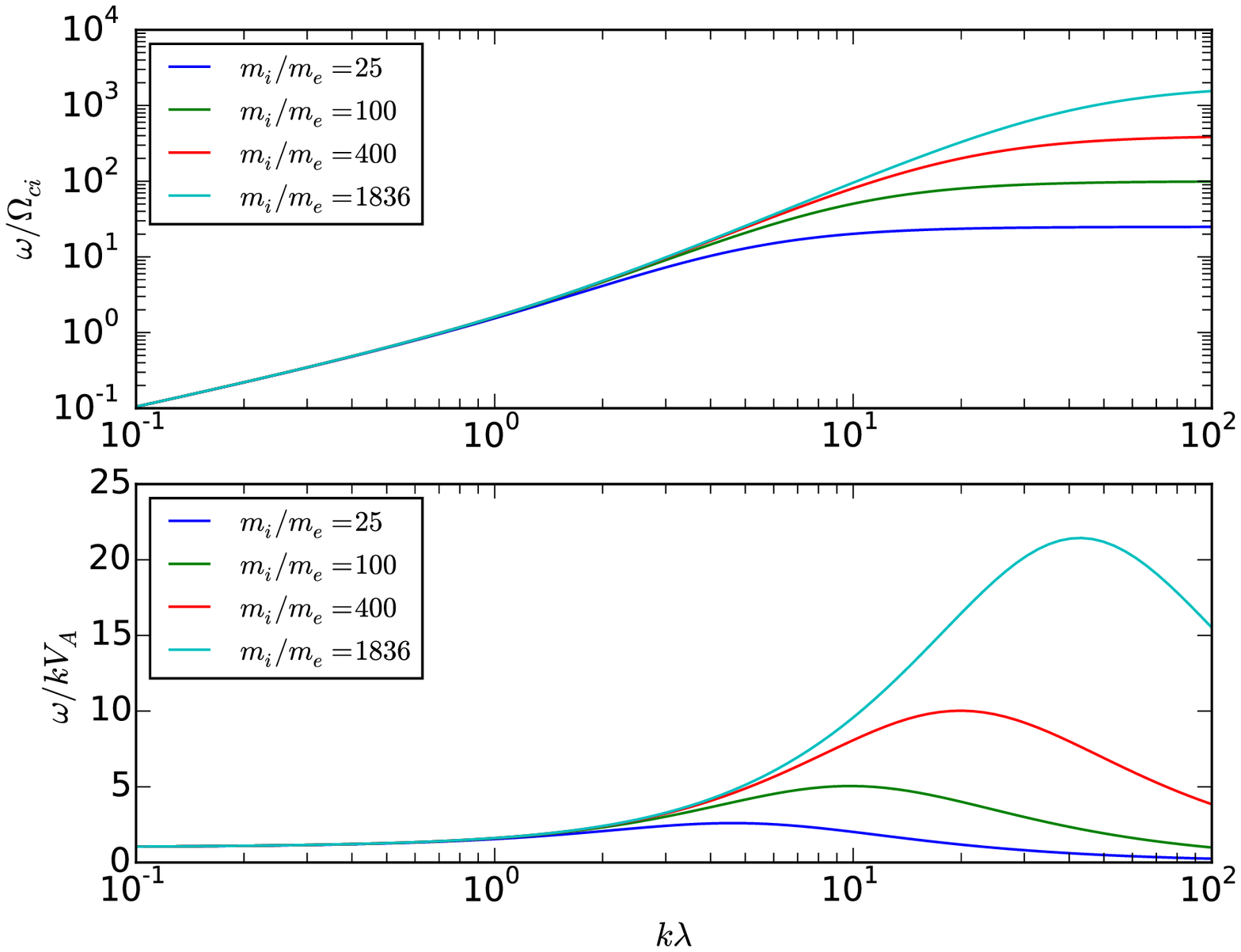}

  \caption{Linear dispersion relation for electromagnetic waves propagating
  parallel to the ambient magnetic field. Wave frequencies as functions of
  wavenumber are shown in the top panel for four different mass ratios
  $m_i/m_e = 25, 100, 400, 1836$. Corresponding phase speeds are shown in the
  bottom panel. Only the waves with maximum phase speed (i.e., on the whistler
  branch) are shown.}

  \label{fig:lindisp}
 \end{center}
\end{figure}

Since the highest phase speed appears at the parallel propagation, we here
investigate this special case in detail. For the parallel propagation, the
sound mode decouples from the electromagnetic modes, and the dispersion
relation may be factorized as follows
\begin{align}
 (X^2 - \beta)
 \left\{
 (1 + \varepsilon \kappa^2)^2 X^4 -
 (2 + (1 + \varepsilon^2) \kappa^2) X^2 +
 1
 \right\}
 = 0.
\end{align}
The dispersion relation for electromagnetic waves corresponding to the second
factor (i.e., the curly bracket) is shown in Fig.~\ref{fig:lindisp}. It is
clearly seen that the wave frequency approaches to the electron cyclotron
frequency at the short wavelength limit. The phase speed, on the other hand,
has a maximum at around $\kappa \simeq \sqrt{m_i/m_e}$ (i.e., $k c/\omega_{pe}
\simeq 1$) and the maximum phase speed is given approximately by $X \simeq
\sqrt{m_i/m_e}/2$.

Note that, in a previous publication, we have shown that the deviation from
the Hall-MHD dispersion $\omega \propto k^2$ occurs approximately at $\kappa
\simeq (m_i/m_e)^{1/4}$ \cite[][Appendix, B.]{2014JCoPh.275..197A}. The weak
dependence of the critical wavenumber on the mass ratio indicates that, at
scale length on the order of or larger than the ion inertial length, the
result should not depend strongly on the reduced mass ratio.

%% References


\bibliographystyle{model5-names} \bibliography{reference}
%\input{ms.bbl}



%\begin{thebibliography}{00}



% \end{thebibliography}
\end{document}